%% file: main.tex
\documentclass[manuscript]{acmart}
\usepackage{graphicx} 
\usepackage{url} 
\usepackage{xspace} 
\usepackage{xcolor} 
\usepackage[table]{xcolor} 
\usepackage{listings} 
\usepackage{hyperref} 
\usepackage{footnote}
\usepackage{multirow} 
\usepackage{booktabs} 
\usepackage{tcolorbox} 
\usepackage{enumitem} 
\lstdefinestyle{prompt}{
    basicstyle=\rm\small,
    breaklines=true,
    tabsize=6, 
    columns=fullflexible,
    frame=single,
    numbers=none,
}
\usepackage{rotating}

\usepackage{pdflscape}
\usepackage{framed}
\usepackage{mdframed}
\usepackage{seqsplit}

\definecolor{niceorange}{rgb}{1.0, 0.6, 0.2}
\definecolor{c0}{rgb}{0.1216, 0.4667, 0.7059}
\colorlet{softc0}{c0!70!white}

\newcommand{\gptfouromini}{\texttt{GPT-4o mini}\xspace}
\newcommand{\geminitwothinking}{Gemini 2.0 Flash Thinking\xspace}
\newcommand{\geminitwofivepro}{Gemini 2.5 Pro\xspace}

\newcommand{\HF}{Hugging Face\xspace}
\newcommand{\reorganizer}{\textsc{MCTidy}\xspace}
\newcommand{\generator}{\textsc{MCGenie}\xspace}

\newcommand{\reorganizationrqcorrect}{How correct are the reorganizations performed by \reorganizer?}
\newcommand{\reorganizationrqhallucination}{How much hallucinated content is introduced by \reorganizer?}
\newcommand{\reorganizationrqconsistent}{How consistent is \reorganizer across multiple runs?}

\newcommand{\generationrqcorrect}{How correct is the information in the generated model cards?}
\newcommand{\generationrqresemble}{How semantically similar are the generated contents to the original model cards?}
\newcommand{\generationrqvaryinginput}{Which input data is the most important for generating model cards?}

\title{Automatic Model Card Generation Using an LLM}
\author{Tajkia Rahman Toma}
\email{tajkiatoma@ualberta.ca}
\affiliation{%
  \institution{University of Alberta}
  \city{Edmonton}
  \country{Canada}
}

\author{Balreet Grewal}
\email{balreet@ualberta.ca}
\affiliation{%
  \institution{University of Alberta}
  \city{Edmonton}
  \country{Canada}
}

\author{Cor-Paul Bezemer}
\email{bezemer@ualberta.ca}
\affiliation{%
  \institution{University of Alberta}
  \city{Edmonton}
  \country{Canada}
}

\begin{document}

\begin{abstract}

Model cards are structured documents that summarize key information about machine learning models to improve transparency, usability, and accountability. However, they often lack a consistent structure, and many models provide no model cards, making comparison and interpretation difficult. This paper presents two contributions. First, we propose \reorganizer, an LLM-based approach that reorganizes existing model cards into a standardized template to improve clarity and comparability. Second, we introduce \generator, an LLM-based system that generates model cards directly from model repository data. We apply \reorganizer to 48 Hugging Face model cards and evaluate information retention, section alignment, hallucination, and stability. Our findings show high information retention with minimal textual loss, accurate section assignment, rare hallucinations primarily in descriptive sections, and strong stability across runs. We assess \generator by generating model cards for the same 48 models and assessing semantic similarity, factual correctness, and sensitivity to input resources. The generated model cards achieved high semantic similarity (mean $\approx$ 0.9); over half were fully correct, and most remaining errors were minor. Generation quality depended strongly on the availability of supporting resources, particularly associated papers. Overall, our findings demonstrate the potential of LLM-based methods to enable scalable, standardized model card documentation.
\end{abstract}



\maketitle

\input{introduction}
\input{study_setup}
\input{result-reorganization}
\input{result-generation}

\input{threats_to_validity}
\input{related_work}

\input{conclusion}

\bibliography{bibtex}
\bibliographystyle{plain}

\input{appendix}

\end{document}

%% file: introduction.tex
\section{Introduction}
Machine learning (ML) models are increasingly deployed in high-stakes domains such as healthcare~\cite{habehh2021machine, javaid2022significance}, finance~\cite{goodell2021artificial, ahmed2022artificial}, and criminal justice~\cite{zavrvsnik2021algorithmic, travaini2022machine}. Model documentation plays a crucial role in providing developers and users with essential information about the development, limitations, and appropriate use of these ML models~\cite{mitchell2019model, richards2020methodology}. Effective documentation enhances model transparency~\cite{mitchell2019model}, which in turn supports accountability~\cite{zainyte2021challenges} and enables users to compare, select, and apply models more effectively~\cite{taraghi2024deep}. In contrast, poor documentation, or the absence of documentation, can lead to misuse, misunderstanding, and frustration among users who need to understand a model's behavior, requirements, and performance in order to make informed decisions~\cite{bhat2023aspirations}. 

Model cards~\cite{mitchell2019model} have emerged as a widely endorsed practice for documenting ML models~\cite{bhat2023aspirations, oreamuno2024state, liang2024s, nunes2024using}. However, many models either lack documentation entirely or are insufficiently documented~\cite{bhat2023aspirations, liang2024s}, increasing the risk of model misuse~\cite{bhat2023aspirations}. Despite its importance, creating and maintaining comprehensive, standardized documentation remains a significant challenge~\cite{huggingfaceAppendix, latendresse2024exploratory}. Writing model documentation from scratch is time-consuming, cognitively demanding, and often not prioritized by model developers, especially in fast-paced or resource-constrained environments~\cite{bhat2023aspirations, latendresse2024exploratory}. Automating the generation of model documentation not only ensures that more model providers provide essential information about the models, but also  encourages broader adoption of standardized documentation practices.

Many recent studies have proposed solutions to support automatic model documentation generation~\cite{fang2020introducing, bhat2023aspirations, richards2020methodology, tsay2020aimmx, crisan2022interactive}. These approaches, however, operate in constrained settings, such as relying on data stored in ML Metadata~\cite{mlmetadata}, descriptions available in computational notebooks, or focusing on alternative documentation formats. In contrast, we propose a solution that leverages resources that are often naturally generated during the model development process, such as model repository files or academic papers, reducing the need for additional structured inputs or specialized environments. By leveraging a large language model (LLM) to produce structured documentation based on such available resources, we can enhance the coverage, consistency, and accessibility of model information at scale.

Our study makes two contributions that build on each other. As our first contribution, we introduce \reorganizer, an approach for automatically reorganizing the content of existing model cards using an LLM to align them with a standard model card template. The goal is to improve the structure and readability of key information and enhance consistency across model documentation, thereby facilitating broader adoption of best practices and enabling more effective comparison and evaluation of models. Using off-the-shelf LLMs such as \geminitwothinking, \reorganizer allows model authors to reorganize their documentation easily, without the need for custom infrastructure or extensive manual edits. We reorganized the content of 48 existing \HF model cards and conducted a systematic evaluation of \reorganizer along three research questions:
\begin{itemize}
    \item \textbf{RQ1.1: \reorganizationrqcorrect} Due to the generative nature of LLMs, evaluating content retention after the reorganization and correct information placement are essential to ensure the integrity of reorganized model cards. \reorganizer retains most content during the reorganization: a median of only 4.5\% of information checklist items derived from the original model cards were missing in the reorganized versions, with minimal textual omission. \reorganizer also placed content correctly in nearly all cases, with only 1.9\% of the (sub)sections containing misplaced content. These findings suggest that \reorganizer performs well in both retaining information and assigning it to appropriate sections, making it a reliable tool for automated model card standardization.
    \item \textbf{RQ1.2: \reorganizationrqhallucination} Since hallucinations and misinterpretations are known limitations of LLMs, it is essential to assess their impact on content reorganization to determine the level of trust we can place in \reorganizer. The overall information accuracy remained high, with only minimal textual changes required to correct the hallucinations and misinterpretations. The hallucinations and misinterpretations were concentrated in specific (sub)sections, suggesting that minimal human oversight focused on these sections is sufficient to catch occasional errors and ensure the overall information accuracy of the reorganized model cards.
    \item \textbf{RQ1.3: \reorganizationrqconsistent} Given the inherently non-deterministic nature of LLMs, we evaluate the consistency of \reorganizer{}'s outputs to assess the stability of its content reorganization across multiple runs. The results demonstrate high semantic consistency, with the median average semantic similarity scores across three runs reaching 0.97. Additionally, 87.5\% of (sub)sections achieved semantic similarity scores of 0.90 or higher, further supporting \reorganizer{}'s suitability for dependable large-scale documentation restructuring.
\end{itemize}

We released both \reorganizer and the 48 reorganized, manually verified model cards as part of our replication package~\cite{replication-package}. These verified reorganized model cards not only serve as a benchmark for future research on automated model card generation but also enable scalable, section-by-section analysis of current documentation practices.

Building on this benchmark, our second contribution explores how LLMs can assist in automatically generating a model card from model repository files. While reorganization ensures structural consistency for existing model cards, many models still lack documentation altogether. In such cases, generation becomes essential. LLMs, with their ability to process and produce text reminiscent of human writing, could potentially automate and streamline the creation of comprehensive, standardized documentation. This would reduce the burden on developers, improve accessibility for users, and encourage more effective use of ML models. To evaluate this potential, we use our reorganized and validated dataset as the ground truth for assessing our approach for automatically generating model cards (\generator), focusing on the following research questions:
\begin{itemize}
    \item \textbf{RQ2.1: \generationrqresemble} We examined how semantically similar the generated model cards are to the original ones, in order to assess \generator{}'s ability to accurately capture and convey the key information from repository files. The generated model cards showed high overall semantic similarity (mean $=$ 0.9), indicating strong information retention. However, sections demanding interpretation or reasoning (such as \texttt{Caveats and Recommendations}) had lower similarity, suggesting that while \generator effectively preserves factual content, it is less consistent in replicating interpretive or context-dependent text.
    \item \textbf{RQ2.2: \generationrqcorrect} We assessed whether the information generated by \generator is factually correct, as LLMs may introduce speculative or unsupported details. Over half (54.17\%) of the generated model cards were entirely correct, and from the remainder, a median generated model card contained only one inaccurate (sub)section. These inaccuracies mainly stemmed from citation-content mismatches or speculative reasoning, suggesting that while \generator reliably reproduces factual content, improving source alignment and grounding could further enhance its factual accuracy.
    \item \textbf{RQ2.3: \generationrqvaryinginput} We examined how variations in input data influenced the generated model cards to understand which repository resources most strongly affect content quality. Repository papers had the greatest impact while configuration or tokenizer files had minimal effect. This implies that papers serve as the primary, information-rich source for \generator, whereas other technical files provide supplementary but non-essential details.
\end{itemize}



The remainder of the paper is structured as follows. Section~\ref{sec:study_setup} discusses our study setup. Sections~\ref{sec:result_reorganization} and~\ref{sec:result_generation} discuss the results and implications of our study on \reorganizer and \generator, respectively. Section~\ref{sec:threats_to_validity} discusses the threats to the validity of this work, and Section~\ref{sec:related_work} summarizes the related work. Section~\ref{sec:conclusion} concludes our work.

%% file: study_setup.tex
\begin{figure}
    \centering
    \includegraphics[width=\columnwidth]{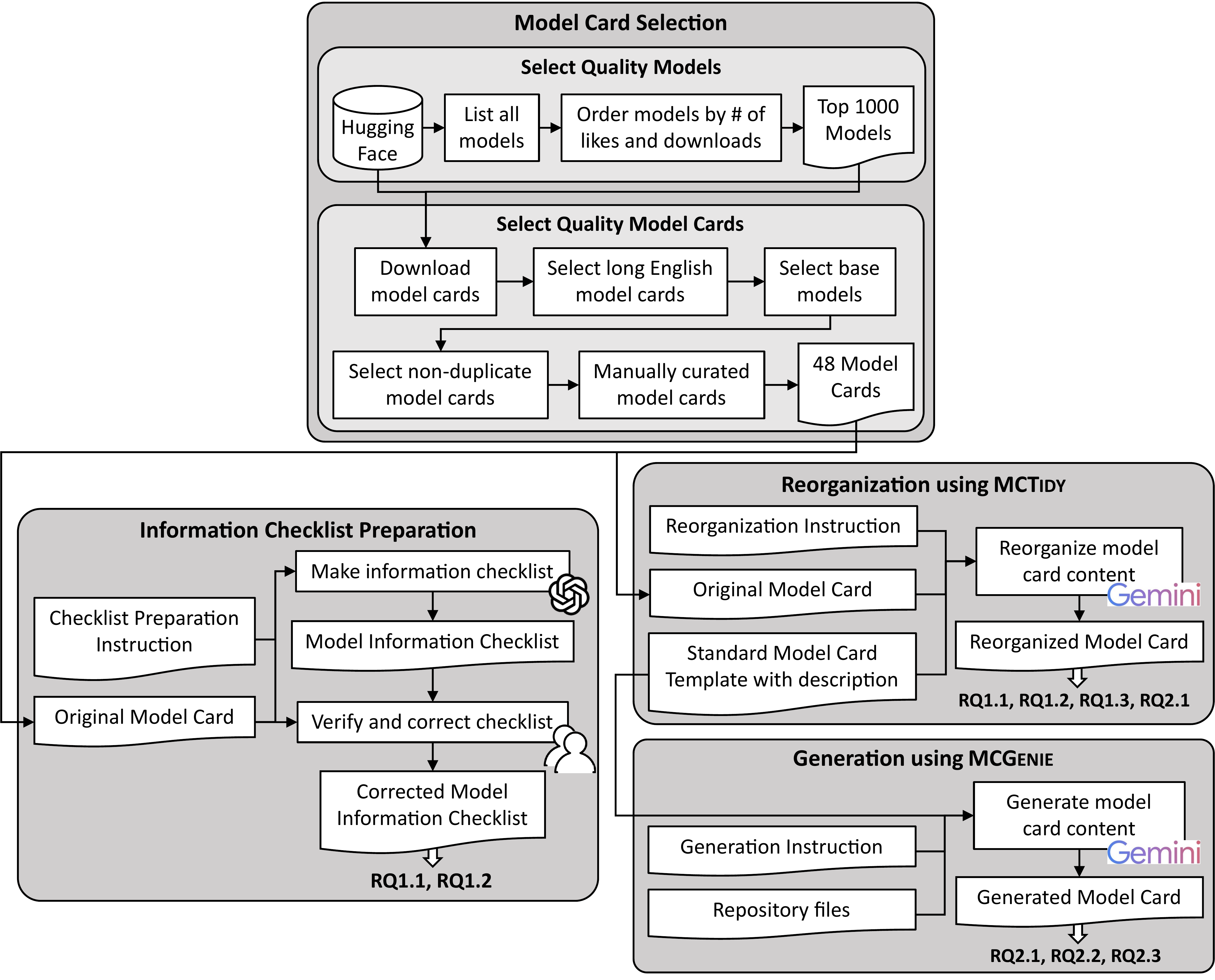}
    \caption{Overview of our study methodology}
    \label{fig:study_setup_overview}
\end{figure}

\section{Study Setup}
\label{sec:study_setup}

We reorganize the content of model cards according to a standard template by using an off-the-shelf LLM. We chose the model card template proposed by Mitchell et al.~\cite{mitchell2019model} as a standard due to its large adoption in recent studies~\cite{bhat2023aspirations, oreamuno2024state, liang2024s, nunes2024using, toma2025answering}. We provided the LLM with both the original model card and the standard template, along with instructions on how to reorganize the content. The template includes descriptions for each section and subsection to guide the LLM during reorganization. The descriptions are based on the framework outlined by Mitchell et al.~\cite{mitchell2019model}. We further refined the descriptions using insights from Toma et al.~\cite{toma2025answering}, and included the two additional sections proposed in their study: ``\texttt{How to Use}'' and ``\texttt{Memory and Hardware Requirements}''. The description of the sections of the model card template is present in~\autoref{sec:appendix}.

Similar to model card reorganization, we generate a model card that follows the same standard template used for reorganization, using an off-the-shelf LLM. However, instead of reorganizing an existing model card, we provided the model with general documentation and resources about the model, such as the associated paper of the model, the tokenizer information saved in JSON file format, configuration files, etc. 
For the section and subsection descriptions in the template, we reused the descriptions defined during the model card reorganization. We instructed the LLM to structure the information from the provided resources into the appropriate sections and subsections of a model card in accordance with the given template. 

In this section, we describe the collection of quality model cards from \HF used for two purposes: reorganizing existing model cards to conform to a standardized template (\reorganizer) to evaluate the effectiveness of the reorganization process, and generating new model cards from model repository data (\generator) to assess the quality of the generation. We then explain the data preparation steps undertaken to facilitate the evaluation of \reorganizer. Finally, we detail the workflows of both \reorganizer and \generator. 
~\autoref{fig:study_setup_overview} provides an overview of the overall process, which is elaborated upon in the remainder of this section. 

\subsection{Model Card Selection}
Although our reorganization approach can be applied to any model card, we select quality models and their corresponding model cards to effectively evaluate whether the reorganization process can handle long and content-rich documentation. These same model cards were also used to evaluate the generation approach, allowing us to compare the generated model cards with human-written ones in terms of resemblance.

\paragraph{Select Quality Models:}
We used the \HF Hub API\footnote{\url{https://huggingface.co/docs/huggingface_hub/package_reference/hf_api\#huggingface_hub.HfApi.list_models}} to collect a list of 944,178 models from the \HF Hub. We then sorted them by the number of likes (indicating overall popularity) and by downloads in the last month (showing current popularity). Then we selected the top 1,000 models, assuming that their model cards are well-maintained.

\paragraph{Select Quality Model Cards:}
We downloaded the model cards of the top 1,000 models for further processing. Since \HF renders \texttt{README.md} files as model cards\footnote{\url{https://huggingface.co/docs/hub/en/model-cards\#model-card-metadata}}, we retrieved the \texttt{README.md} files from the repositories using the \texttt{huggingface\_hub} library\footnote{\url{https://huggingface.co/docs/huggingface\_hub/en/guides/download}}. We found 993 repositories with model cards. Then, we filtered them for quality using the following selection criteria:

\begin{itemize}
    \item \textbf{English model cards:} To ensure consistent understanding of the content, we chose to include model cards written only in English, the most commonly used language. We filtered out non-English model cards using \texttt{xlm-roberta-base-language-detection}~\cite{luca_papariello_2024}, leaving us with 926 model cards.
    \item \textbf{Long model cards:} We selected model cards that are sufficiently detailed, retaining those with at least 1,000 words of descriptive content, excluding code blocks from the word count. Since code blocks focus on how to use the model or output of the model, rather than describing the model, a large code block can make a model card long without adding enough details about the model itself (e.g., the model card of \texttt{MEETING\_SUMMARY}\footnote{\url{https://huggingface.co/knkarthick/MEETING_SUMMARY}}). This step resulted in 231 model cards. 
    \item \textbf{Model cards of base models:} Model cards for derivative models (adapters, fine-tunes, merges, and quantizations) often extend the base model's card. For example, the model cards of \texttt{Llama-2-7b-chat-hf}\footnote{\url{https://huggingface.co/meta-llama/Llama-2-7b-chat-hf}} and \texttt{Llama-2-7B-Chat-GGML}\footnote{\url{https://huggingface.co/TheBloke/Llama-2-7B-Chat-GGML}} show how model cards of quantized models combine content from both the base model and the quantized model. Therefore, to focus on original content, we selected only base models and excluded all model cards of derivative models. Whether a model is a base or derivative is indicated in the model card's YAML metadata. After this filtering step, we were left with 176 model cards of base models.
    \item \textbf{Non-duplicate model cards:} Model cards from the same model family (e.g., fine-tuned from the same base model with different numbers of parameters) are often quite similar (e.g., the model card of \texttt{gemma-7b}\footnote{\url{https://huggingface.co/google/gemma-7b}} and \texttt{gemma-7b-it}\footnote{\url{https://huggingface.co/google/gemma-7b-it}}). Although we filtered for base models in the previous step, not all model cards clearly specify their base model, and models from the same family are often from the same organization. Additionally, model cards from the same organization often follow a consistent structure and writing style, even across different models (e.g., the model card of \texttt{Llama-2-7b}\footnote{\url{https://huggingface.co/meta-llama/Llama-2-7b}} and \texttt{Meta-Llama-3-8B}\footnote{\url{https://huggingface.co/meta-llama/Meta-Llama-3-8B}}). Therefore, to mitigate potential bias from structurally similar model cards, we selected only one model (the top one) from each organization. This left us with 61 model cards.
    \item \textbf{Manually curated model cards:} One author manually removed repositories that are not model repositories (e.g., the \texttt{hollowstrawberry/stable-diffusion-guide}\footnote{\url{https://huggingface.co/hollowstrawberry/stable-diffusion-guide}} repository), do not contain a model card for a single model (e.g., model card of \texttt{YoungMasterFromSect/Trauter\_LoRAs}\footnote{\url{https://huggingface.co/YoungMasterFromSect/Trauter_LoRAs}}), or were incorrectly included due to automation errors (e.g., model card of \texttt{CausalLM/14B}\footnote{\url{https://huggingface.co/CausalLM/14B}}, which contains non-English text). After this final step, we were left with 48 model cards.
\end{itemize}

\subsection{Information Checklist Preparation}
\label{sec:checklist_preparation}

When reorganizing a model card, sentences from the original model card may be split and redistributed across different sections. This fragmentation makes it difficult to verify whether all information from the original model card is preserved in the reorganized version. To facilitate systematic comparison, we introduce an intermediate step that constructs a checklist from the original model card, decomposing each sentence into distinct units of information. For example, we break down the following sentence:

\smallskip
\noindent
\begingroup
\leftskip=1em
\rightskip=1em
\itshape
Stable Diffusion is a latent text-to-image diffusion model capable of generating photo-realistic images given any text input.
\par
\endgroup
\smallskip

into these three sentences:

\begin{enumerate}[leftmargin=*]
    \item \textit{Stable Diffusion is a latent text-to-image diffusion model.}
    \item \textit{Stable Diffusion is capable of generating photo-realistic images.}
    \item \textit{Stable Diffusion works with any text input.}
\end{enumerate}



After the reorganization of model card contents, the following types of information discrepancies may arise:
\begin{itemize}
    \item \textbf{Missing information:} Information included in the original model card but absent in the reorganized model card.
    \item \textbf{Extra information:} Information included in the reorganized model card but absent from the original model card, typically due to hallucination or misinterpretation. While such statements may seem plausible, they are considered extraneous since they were not explicitly present in the original.
\end{itemize}

To identify such discrepancies in the reorganized model cards, we used these checklists. By breaking down the original model card at the information level, the checklists enabled a more granular and systematic verification of content preservation in the reorganized versions. However, manually creating the checklists is time-consuming and error-prone. Given the strength of LLMs to understand and process natural language, we used \texttt{gpt-4o-mini-2024-07-18} to generate them. The instructions for checklist preparation can be found in our replication package. To verify the checklists, two authors of this paper manually reviewed the checklists in three rounds, identifying and correcting missing or extra information in the checklists. After each round, they compared their results and resolved disagreements through discussion. 

A median checklist had 69 items. We computed the normalized Levenshtein distance between the checklists generated by \gptfouromini and their manually corrected counterparts to evaluate the model's performance in checklist generation. This metric quantifies the extent of edits required to correct the generated checklists. The average normalized Levenshtein distance between the \gptfouromini-generated and manually corrected checklists is $0.03$. This indicates that the generated checklists are very close to the manually revised versions, requiring only minimal edits. Such a low distance reflects strong alignment with human expectations, suggesting that the end users of \reorganizer can use the checklists directly for their verification too. The usage of the checklists to identify missing (RQ1.1) and extra (RQ1.2) information is detailed in the respective sections.

\subsection{Reorganization Using \reorganizer}
\label{sec:reorganization_using_mctidy}
Listing \ref{list:reorganization_prompt_template} presents the prompt we used to reorganize the content of the 48 model cards with \texttt{\seqsplit{gemini-2.0-flash-thinking-exp-01-21}}, using a temperature setting of $0$. We provide a standard model card template along with the original model card in the prompt to guide the reorganization. The template specifies the required sections and subsections, along with descriptions of the expected content for each. The 48 original model cards and the standard model card template, including section and subsection descriptions, are available in our replication package~\cite{replication-package}. 

\begin{figure}[tb]
    \centering
\begin{lstlisting}[language={}, style=prompt]
You are an AI assistant tasked with reorganizing model card content to fit a provided template. Your goal is to place existing information into the correct sections and subsections without altering or adding any new content.

**Instructions:**

1. Carefully review the provided model card content and the model card template.
2. For each section and subsection in the template, locate the corresponding information within the model card content.
3. Move the information to the appropriate section and subsection in the template. Merge scattered information into the correct sections of the template. If a section already exists, you can move relevant details from other parts to complete that section.
4. If a section or subsection in the template does not have corresponding details in the model card content, write "Not available." in that section or subsection.
5. If the model card content includes information (even part of a sentence) that does not fit into any section or subsection of the template, create a new section titled "Additional Information" at the end of the reorganized model card. Place all extra information under this section.
6. Ensure every piece of content (including every details, explanations, reasoning, examples, images, tables, code blocks, citation, links, and emojis) from the original model card is present and that no new content is added. Your task is solely to reorganize the existing content.

**Model Card Content:** """<the model card>"""

**Model Card Template:**"""<the model card template>"""

**Reorganized Model Card:**
\end{lstlisting}
    \caption{Prompt template used to reorganize a model card's content. The model card template with the section descriptions is available in our replication package.}
    \label{list:reorganization_prompt_template}
\end{figure}

\subsection{Generation Using \generator}
To generate model cards, we first removed the existing model cards from the 48 model repositories to avoid influencing the generation process. The model repository files were then provided to \texttt{gemini-2.5-pro} along with detailed generation instructions and the standard model card template prepared in Section~\ref{sec:reorganization_using_mctidy}. \generator was configured with a temperature setting of $0$ to ensure deterministic outputs. The full prompt used for generation is presented in Listing~\ref{list:generation_prompt_template}.

\begin{figure}[tb]
    \centering
\begin{lstlisting}[language={}, style=prompt]
System Instruction:
A model card is a document designed to provide clear, accessible information about a machine learning model. It helps its users and stakeholders understand how the model works, what data it was trained on, how it should be applied etc.

You will be provided with a model repository. The repository may include various materials such as model development code, academic paper describing the model, and other related resources. Your task is to **write a model card using only the information available in the provided repository**. 

All content in the model card must be based on the provided repository data, and **every piece of information used must be accompanied by a citation to its source**. If the provided data do not contain sufficient information for a particular section or subsection, write **"Insufficient information"** for that section or subsection.

Below is the template of a standard model card enclosed in triple quotes. The template contains **11 sections**, some of which include multiple subsections. You need to fill in all sections and subsections comprehensively based on the available information. Each section and subsection contains instructions describing what content to include - replace these instructions with the actual content derived from the repository.
"""<the model card template>"""

-----------------------------------------------
User Message:
Write down the model card from the model's attached repository files. Below are the names of the attached repository files:
"""<the model repository file names>"""
\end{lstlisting}
    \caption{Prompt template used to generate a model card from its content. The model card template with the section descriptions is available in our replication package.}
    \label{list:generation_prompt_template}
\end{figure}

To prepare the model repository files for input to \generator, we cloned the \HF repository of each model, excluding files stored in Git Large File Storage (LFS) and listed the files. Since the \texttt{\geminitwofivepro} model does not support files with \texttt{.svg} and \texttt{.gif} extensions, we removed those from the list. We further excluded files that were not informative for model card generation, such as model parameter files, serialized or sharded model checkpoints, and other large binary artifacts. Finally, we downloaded the associated academic or white papers available for each repository and included them in the list. 

Then, we provided the listed model repository files for each model to \texttt{\geminitwofivepro} for corresponding model card generation. We successfully generated 26 model cards without exceeding the maximum input token limit of \texttt{\geminitwofivepro}. For the remaining 22 repositories that produced errors, we examined common sources of large input files and found that many contained a large \texttt{tokenizer.json} file. Since a \texttt{tokenizer.json} file is informative for model card generation, we created summarized versions of it and resubmitted the inputs to \generator. We summarize each tokenizer JSON file by extracting its model type, vocabulary size, and small samples of vocabulary entries and merge rules. We also retain the normalization, pre-tokenization, post-processing, and decoding configurations, along with all explicitly marked special tokens. This produces a compact representation that preserves essential tokenizer characteristics while reducing file size.

Additionally, two repositories, \texttt{deepseek-ai/DeepSeek-V2-Chat} and \texttt{facebook/seamless-m4t-v2-large}, included large \texttt{model.safetensors.index.json} and \texttt{generation\_config.json} files, respectively. We similarly summarized key information from these files for inclusion. Thus, we could generate 18 more model cards without the error.

Among the remaining 4 model repositories, we found 3 repositories, \texttt{Qwen/Qwen2-VL-7B-Instruct}, \texttt{\seqsplit{Tencent-Hunyuan/HunyuanDiT}}, \texttt{openai/whisper-large-v3}, containing large \texttt{vocab.json} files. Since the \texttt{tokenizer.json} file partially includes the information from \texttt{vocab.json}, we removed these vocabulary files from the file list. In the last repository, \texttt{ctheodoris/Geneformer}, we identified a large notebook, containing example usage, that significantly increased the total input token count. Although this resulted in some information loss, the notebook was removed to fit within the input constraints.


%% file: result-reorganization.tex
\section{Evaluation of \reorganizer}
\label{sec:result_reorganization}
In this section, we present the evaluation of our LLM-based model card reorganization approach, \reorganizer.

\input{result-reorganization-correct}
\input{result-reorganization-accuracy}
\input{result-reorganization-consistency}
\input{discussion-reorganization}

%% file: result-reorganization-correct.tex
\subsection{RQ1.1: \reorganizationrqcorrect}
\label{sec:reorganization_rq_correct}

\textit{Motivation:}
Since LLMs may omit or misplace information due to their generative nature, it is important to evaluate (1)~how well the reorganized model cards retain all key content from the original model card and (2)~whether it is assigned to the appropriate sections.

\textit{Approach:}
To evaluate information retention, we used the corrected model information checklists from Section~\ref{sec:checklist_preparation}. One author manually reviewed each reorganized model card alongside its checklist, marking each item in the checklist as present, partially missing, or missing. All missing items and the missing portions of partially missing items were then manually added to the ``Additional Information'' section of the reorganized model card to complete it with the full set of information.

We quantified information retention by calculating the percentage of present checklist items per model card. We also computed the Levenshtein distance between each reorganized model card and its corresponding manually completed version, normalized by the length of the completed version. While the percentage of present checklist items quantifies the proportion of documentation information that was captured during the reorganization process, the normalized Levenshtein distance captures the extent of information relative to the full documentation that was not captured.

We also assessed whether longer model cards were more prone to omissions. Therefore, we calculated the Pearson correlation between the number of total checklist items and the number of missing items. We interpreted the strength of the correlation using commonly accepted thresholds~\cite{schober2018correlation}: 0.00--0.10 (negligible), 0.10--0.39 (weak), 0.40--0.69 (moderate), 0.70--0.89 (strong), and 0.90--1.00 (very strong). Additionally, we computed the p-value to assess the statistical significance of the correlation, with p $<$ 0.05 indicating a significant relationship.

To evaluate whether checklist items were placed into the appropriate model card sections, we used an LLM jury~\cite{verga2024replacing}, an increasingly popular method~\cite{li2024software, kenton2024scalable, seo2025large} that simulates multiple independent evaluators and aggregates their judgments (e.g., through majority voting) to reduce bias and improve robustness. We employed three LLMs: \texttt{Gemini-2.5-Pro}, \texttt{OpenAI o4-mini}, and \texttt{DeepSeek R1} as jury models, selected based on cost analysis and their rankings on the Arena leaderboard~\cite{chiang2024chatbot}. We asked each juror model to answer ``Is the content in this section relevant to this section?'' as ``yes'' or ``no''. If judged irrelevant, the model was also asked to identify the irrelevant parts.

We used majority voting across the LLMs to determine whether a section contained misplaced content. Two authors then manually reviewed each flagged section to validate the models’ judgments and confirmed whether the identified content was indeed misplaced based on their judgment. We quantified section appropriateness by calculating the percentage of (sub)sections that contained confirmed misplaced content out of all (sub)sections in the 48 reorganized model cards. This reflects how well \reorganizer aligned content with appropriate sections.

\begin{figure}[tb]
    \centering
    \includegraphics[width=.6\columnwidth]{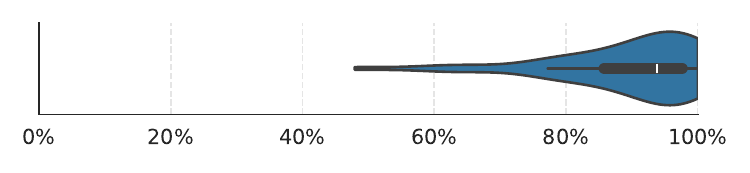}
    \caption{Distribution of retained checklist items (in percentage) across 48 model cards}
    \label{fig:present_points_percentage_distribution}
\end{figure}

\textit{Findings:}
\textbf{\reorganizer retained a median of 93.8\% of the information from the original model card to the reorganized model card.} ~\autoref{fig:present_points_percentage_distribution} shows that the best-performing model cards (6) retained all items, while the worst one retained 47.9\% of items. The median reorganized model card had only 1.6\% of checklist items partially missing and only 4.5\% completely missing. We also investigated whether model cards with more checklist items were likely to have more missing items, but the Pearson correlation between the number of checklist items and the number of missing items was not statistically significant (p = 0.1310). 

\begin{figure}[tb]
    \centering
    \includegraphics[width=.6\linewidth]{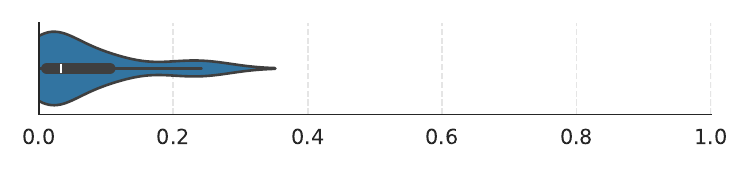}
    \caption{Distribution of normalized Levenshtein distances between the 48 reorganized model cards and their corresponding manually \textit{completed} versions to measure the proportion of information from the full model card that was not captured}
    \label{fig:completeness_edit_distance_distribution}
\end{figure}

\textbf{Only a small number of edits were needed to manually complete most of the reorganized model cards.} As shown in~\autoref{fig:completeness_edit_distance_distribution}, the median normalized Levenshtein distance between the reorganized and manually completed cards was 0.03, indicating very low textual deviation. The best-performing reorganizations that retained all checklist items had a normalized distance of 0.0 (no edits were needed). In contrast, the worst-performing reorganization had a normalized distance of 0.28, reflecting substantial missing content relative to the full model card. 

Upon closer inspection of the partially and completely missing checklist items, \textbf{we observed several patterns of common omissions}. Some omissions were acceptable due to redundancy or low relevance, while others involved important content that ideally should have been retained. The most frequently missing types of content included:
\begin{itemize}[leftmargin=*]
    \item \textbf{Practical usage guidance}: These are tips or examples to help users interact with the model, such as how to format inputs or understand outputs. For example, the model card for \texttt{CAMB-AI/MARS5-TTS} includes tips like adding commas to insert pauses in speech synthesis, and similar formatting guidance appears in \texttt{jbetker/tortoise-tts-v2} and \texttt{CohereForAI/c4ai-command-r-plus}. These tips are important for the model's end user and should be retained in the reorganized version.
    
    \item \textbf{Explanatory statements}: These are explanations or descriptions of why certain tasks were done. For instance, the sentence \textit{``Our goal in performing this evaluation was to try to identify ...''} from \texttt{HuggingFaceM4/idefics2-8b} was omitted. While not directly related to the model's behavior, such explanations can provide important context and enhance the model's transparency, and are therefore worth retaining.
    
    \item \textbf{Subjective content}: These are contents like personal opinions, community feedback, or general statements. For instance, informal phrases like \textit{``You may love or hate it''} (e.g., \texttt{Crosstyan/BPModel}) or general statements like \textit{``Language models are widely used for tasks other than token prediction''} from \texttt{EleutherAI/gpt-j-6b} were excluded. While such content may not be strictly factual or technical, it can sometimes help explain a design choice, provide context, or clarify the intent behind certain decisions. As such, retaining them could improve the readability or interpretability of the model card.
    
    \item \textbf{Introductory blurbs}: These are short phrases in the \texttt{Citation} subsection asking users to cite the model before showing the actual citation. They were often omitted, like in \texttt{ibm-granite/granite-timeseries-ttm-v1} and \texttt{openlm-research/open-llama-13b}). Since they do not provide additional information beyond the citation itself, their omission does not negatively impact the completeness or utility of the reorganized model card.
        
    \item \textbf{Navigational or meta-information phrases}: These are references that help users move through the document but do not add specific model-related details. For example, in \texttt{openai/whisper-large-v3}, the sentence \textit{``For more details on the different checkpoints available, refer to the section [Model details](\#model-details)''} was omitted. Similar references to internal sections were also missing in \texttt{NousResearch/Llama-2-7b-chat-hf}, \texttt{HuggingFaceM4/idefics2-8b}, and \texttt{openchat/openchat\_3.5}. Including such phrases in the reorganized version can sometimes create confusion by referring to incorrect or missing sections in the reorganized version, therefore, it is better to omit them.
\end{itemize}

\textbf{\reorganizer places 98.1\% of the content in the correct section.} Out of the 1,440 (sub)sections across the 48 reorganized model cards, only 28 (1.9\%) contained content that was confirmed to be misplaced in that section. 
\textbf{The misplaced content was distributed across various sections}. As shown in the `\# with misplaced information' column in~\autoref{tab:merged_table}, no single (sub)section consistently suffered from misplacement. The highest number of confirmed misplaced instances (4) appeared in the \texttt{Quantitative Analyses/Unitary results} subsection, followed by four other (sub)sections with three cases each. Given the broad distribution of these instances, we cannot confidently attribute misplacement to any specific (sub)section.

In 10 cases, the LLM jury flagged sections as containing misplaced content, but human reviewers disagreed. The conflict often stemmed from the jury's strict adherence to section descriptions, whereas human evaluators applied a more flexible, context-aware perspective. 
Jurors also occasionally flagged sections as containing misplaced content due to a lack of expected elaboration. 
Human reviewers still marked them as relevant, as the content appropriately belonged to the section. 

\begin{figure}[t!]
    \centering
    \includegraphics[width=.6\columnwidth]{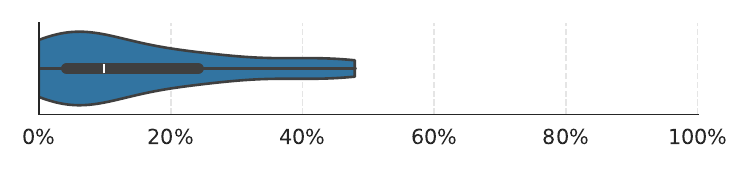}
    \caption{Distribution of the ``Additional Information '' section length as a percentage of the total model card length across 48 model cards}
    \label{fig:additional_info_section_size_ratio_distribution}
\end{figure}

To verify content placement, we also analyzed the amount of information placed in the ``Additional Information'' section. Since this section can contain any type of content without being considered misplaced into that section, we could not apply the same evaluation process we followed for other sections. Instead, we measured the proportion of content it contains relative to the entire document to assess whether it tends to hold a large share of information, which could impact the accuracy and clarity of content placement within the model card. As shown in~\autoref{fig:additional_info_section_size_ratio_distribution}, the ``Additional Information'' section is a median of 9.8\% of the total model card, with 4 model cards not having an additional information section, indicating that this section generally comprises a small portion of the full model card.

\begin{tcolorbox}
\reorganizer demonstrates strong overall performance in retaining key content and assigning it to appropriate sections. It achieves a high median checklist retention rate of 93.8\% and a low median normalized Levenshtein distance of 0.03, indicating minimal textual omissions. While some missing content, such as usage guidance or subjective explanations, is important, other omissions like introductory blurbs or navigational phrases, have limited effect on overall completeness. Content placement was also highly accurate, with only 1.9\% of (sub)sections from all the 48 model cards containing misplaced content, suggesting content misplacement is occasional rather than systematic. Their low frequency and limited impact indicate that minimal human oversight is sufficient to maintain the quality of the reorganized model cards.
\end{tcolorbox}

%% file: result-reorganization-accuracy.tex
\subsection{RQ1.2: \reorganizationrqhallucination}
\label{sec:rq_accuracy}

\textit{Motivation:}
As LLMs hallucinate, there can be some extra information that is not present in the original model card. Therefore, we examined the reorganized model cards for added content to evaluate how much new information was introduced during reorganization to understand how much we can trust \reorganizer.

\textit{Approach:}
Similar to identifying missing information in reorganized model cards, we used the corrected model information checklists to identify extra information. While reviewing each reorganized model card for completeness, if we found content that was not present in the checklist, we marked it as extra information and removed it from the reorganized version. However, if the generated information was a misinterpretation of the original model card, we corrected it. 

Similar to our approach for RQ1.1, we assessed the accuracy of information in the model cards reorganized by \reorganizer by computing the Levenshtein distance between each reorganized model card and its manually corrected version, normalized by the length of the reorganized version. We then reported the distribution of these distances across all model cards. This metric captures the overall textual deviation between the reorganized and corrected versions, reflecting the extent of extra information in proportion to the full documentation.

\begin{figure}[tb]
    \centering
    \includegraphics[width=.6\linewidth]{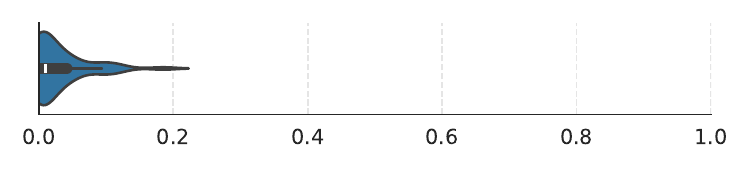}
    \caption{Distribution of normalized Levenshtein distances between the 48 reorganized model cards and their corresponding manually \textit{corrected} versions to measure the extent of hallucinated information relative to the full model card}
    \label{fig:correction_edit_distance_distribution}
\end{figure}

\textit{Findings:}
\textbf{Hallucination was minimal across reorganized model cards.} Most model cards require little to no manual editing to remove or correct hallucinated content. As shown in~\autoref{fig:correction_edit_distance_distribution}, the median normalized Levenshtein distance between a reorganized model card and its corrected version is 0.01, indicating a very low level of textual deviation.  The best-performing reorganizations (11) had a normalized distance of 0.0 (no edits were needed). In contrast, the worst-performing reorganization had a normalized distance of 0.19, reflecting substantial incorrect content generated relative to the full model card. 

\textbf{Most section-level corrections were concentrated in a few specific areas.} We found hallucinations in a total of 147 (10.2\%) (sub)sections out of 1,440 (sub)sections from the 48 model cards. From~\autoref{tab:merged_table} we see that hallucinations occurred most often in the \texttt{Primary intended users} subsection from the \texttt{Intended Use} section and in the \texttt{Ethical Considerations} section. These sections typically contain descriptive or interpretive content, rather than concrete technical details about the model itself. As a result, they may leave more room for ambiguity or inferred content during reorganization.

\textbf{We observed recurring patterns in the types of extra information introduced by \reorganizer}. In the \texttt{Primary intended users} subsection, \reorganizer introduced extra content by generalizing or expanding the list of user groups in almost each of the 15 cases. For example, it frequently added long lists such as ``researchers, developers, businesses, and educators'', even when the original model card mentioned only a few of these roles or none at all. This suggests that \reorganizer tends to assume a broad audience for models.

In the \texttt{Ethical Considerations} section, we observed that \reorganizer frequently (in 10 model cards) included the section description verbatim from the provided template. Some model cards also included general ethical or safety statements that were not grounded in the original content. For example, in the \texttt{DeepSeek-AI/DeepSeek-V2-Chat} model card, \reorganizer added broad claims about safety practices that were not documented in the original model card.

We also found made-up content in the \texttt{Motivation} sections of the model cards, specifically, \texttt{\seqsplit{Training Data/Motivation}} and \texttt{Evaluation Data/Motivation}. In these cases, \reorganizer introduced motivations that were not present in the original model card. These hallucinated motivations were likely inferred from model characteristics (e.g., size or architecture); however, they lacked direct support from the original model card. Additionally, \reorganizer often appended explanatory phrases after inferred content, such as ``This implies ...'' or ``This suggests ...'' (e.g., in \texttt{openchat/openchat-3.5}), regardless of the (sub)section.

We also observed repeated issues with misinterpretations of referential terms such as ``following'', ``below'', or ``these''. \reorganizer often kept these expressions without correctly identifying what they referred to, which led to unclear or misleading references. In some cases, it claimed that certain information could be found in sections that were present in the original model card but did not exist in the reorganized card, or misattributed content to the wrong section. These types of errors created confusion and reduced the usefulness of the model card by pointing readers to content that was not present.




\begin{tcolorbox}
Only 10.1\% of the (sub)sections of the reorganized model cards required minor corrections for extra or misinterpreted content. Hallucinations occurred the most often in the \texttt{Primary intended users} and \texttt{Ethical Considerations} sections, indicating that while accuracy is generally high, it can be helpful to conduct a manual review of certain subsections after reorganizing a model card.
\end{tcolorbox}

\begin{landscape}
\begin{table}[tb]
    \small
    \caption{The number of reorganized model cards with misplaced information (RQ1) or hallucinations (RQ2), and the semantic similarity per subsection (RQ3). For RQ1 and RQ2, darker colors mean worse outcomes (more misplaced information or hallucinations), and for RQ3, darker colors mean better outcomes (higher similarity). }
    \label{tab:merged_table}
    \centering
	\input{fig/reorganization_merged_data_table}
\end{table}
\end{landscape}

%% file: fig/reorganization_merged_data_table.tex
\begin{tabular}{@{}lrrrr@{}}
    \toprule
    \multirow{5}{*}{\textbf{(Sub)sections}} & \multicolumn{1}{c}{\textbf{RQ1.1}} & \multicolumn{1}{c}{\textbf{RQ1.2}} & \multicolumn{2}{c}{\textbf{RQ1.3}} \\
    \cmidrule(lr){2-2} \cmidrule(lr){3-3} \cmidrule(lr){4-5}
     & \multirow{3}{*}{\shortstack{\textbf{\# with} \\ \textbf{misplaced} \\
     \textbf{information} \\ \textbf{(out of 48)}}} & \multirow{3}{*}{\shortstack{\textbf{\# with} \\ \textbf{hallucinations} \\ \textbf{(out of 48)}}} & \multicolumn{2}{c}{\textbf{Semantic}} \\
    & & & \multicolumn{2}{c}{\textbf{similarity scores}} \\
    \cmidrule(lr){4-5}
    & & & \textbf{Avg.} & \textbf{Med.} \\
    \midrule
    	Model Details/Person or organization developing model & \cellcolor{niceorange!2.083333333333333!white}1 (2.1\%) & \cellcolor{niceorange!16.666666666666664!white}8 (16.7\%) & \cellcolor{softc0!94.00!white}0.94 & \cellcolor{softc0!97.00!white}0.97 \\ 
		Model Details/Model date & \cellcolor{niceorange!0.0!white}0 (0.0\%) & \cellcolor{niceorange!10.416666666666668!white}5 (10.4\%) & \cellcolor{softc0!92.00!white}0.92 & \cellcolor{softc0!93.00!white}0.93 \\ 
		Model Details/Model version & \cellcolor{niceorange!0.0!white}0 (0.0\%) & \cellcolor{niceorange!8.333333333333332!white}4 (8.3\%) & \cellcolor{softc0!92.00!white}0.92 & \cellcolor{softc0!92.00!white}0.92 \\ 
		Model Details/Model type & \cellcolor{niceorange!0.0!white}0 (0.0\%) & \cellcolor{niceorange!8.333333333333332!white}4 (8.3\%) & \cellcolor{softc0!95.00!white}0.95 & \cellcolor{softc0!95.00!white}0.95 \\ 
		Model Details/Training details & \cellcolor{niceorange!6.25!white}3 (6.2\%) & \cellcolor{niceorange!4.166666666666666!white}2 (4.2\%) & \cellcolor{softc0!95.00!white}0.95 & \cellcolor{softc0!96.00!white}0.96 \\ 
		Model Details/Paper or other resource for more information & \cellcolor{niceorange!2.083333333333333!white}1 (2.1\%) & \cellcolor{niceorange!10.416666666666668!white}5 (10.4\%) & \cellcolor{softc0!93.00!white}0.93 & \cellcolor{softc0!95.00!white}0.95 \\ 
		Model Details/Citation details & \cellcolor{niceorange!0.0!white}0 (0.0\%) & \cellcolor{niceorange!0.0!white}0 (0.0\%) & \cellcolor{softc0!99.00!white}0.99 & \cellcolor{softc0!100.00!white}1.00 \\ 
		Model Details/License & \cellcolor{niceorange!0.0!white}0 (0.0\%) & \cellcolor{niceorange!8.333333333333332!white}4 (8.3\%) & \cellcolor{softc0!98.00!white}0.98 & \cellcolor{softc0!100.00!white}1.00 \\ 
		Model Details/Contact & \cellcolor{niceorange!0.0!white}0 (0.0\%) & \cellcolor{niceorange!2.083333333333333!white}1 (2.1\%) & \cellcolor{softc0!97.00!white}0.97 & \cellcolor{softc0!100.00!white}1.00 \\ 
		Intended Use/Primary intended uses & \cellcolor{niceorange!4.166666666666666!white}2 (4.2\%) & \cellcolor{niceorange!8.333333333333332!white}4 (8.3\%) & \cellcolor{softc0!96.00!white}0.96 & \cellcolor{softc0!96.00!white}0.96 \\ 
		Intended Use/Primary intended users & \cellcolor{niceorange!0.0!white}0 (0.0\%) & \cellcolor{niceorange!31.25!white}15 (31.2\%) & \cellcolor{softc0!92.00!white}0.92 & \cellcolor{softc0!94.00!white}0.94 \\ 
		Intended Use/Out-of-scope uses & \cellcolor{niceorange!0.0!white}0 (0.0\%) & \cellcolor{niceorange!2.083333333333333!white}1 (2.1\%) & \cellcolor{softc0!96.00!white}0.96 & \cellcolor{softc0!97.00!white}0.97 \\ 
		How to Use & \cellcolor{niceorange!0.0!white}0 (0.0\%) & \cellcolor{niceorange!8.333333333333332!white}4 (8.3\%) & \cellcolor{softc0!97.00!white}0.97 & \cellcolor{softc0!98.00!white}0.98 \\ 
		Factors/Relevant factors & \cellcolor{niceorange!0.0!white}0 (0.0\%) & \cellcolor{niceorange!8.333333333333332!white}4 (8.3\%) & \cellcolor{softc0!89.00!white}0.89 & \cellcolor{softc0!90.00!white}0.90 \\ 
		Factors/Evaluation factors & \cellcolor{niceorange!6.25!white}3 (6.2\%) & \cellcolor{niceorange!6.25!white}3 (6.2\%) & \cellcolor{softc0!93.00!white}0.93 & \cellcolor{softc0!91.00!white}0.91 \\ 
		Metrics/Model performance measures & \cellcolor{niceorange!6.25!white}3 (6.2\%) & \cellcolor{niceorange!10.416666666666668!white}5 (10.4\%) & \cellcolor{softc0!92.00!white}0.92 & \cellcolor{softc0!92.00!white}0.92 \\ 
		Metrics/Decision thresholds & \cellcolor{niceorange!0.0!white}0 (0.0\%) & \cellcolor{niceorange!2.083333333333333!white}1 (2.1\%) & \cellcolor{softc0!99.00!white}0.99 & \cellcolor{softc0!100.00!white}1.00 \\ 
		Metrics/Variation approaches & \cellcolor{niceorange!0.0!white}0 (0.0\%) & \cellcolor{niceorange!6.25!white}3 (6.2\%) & \cellcolor{softc0!95.00!white}0.95 & \cellcolor{softc0!100.00!white}1.00 \\ 
		Evaluation Data/Datasets & \cellcolor{niceorange!2.083333333333333!white}1 (2.1\%) & \cellcolor{niceorange!16.666666666666664!white}8 (16.7\%) & \cellcolor{softc0!93.00!white}0.93 & \cellcolor{softc0!94.00!white}0.94 \\ 
		Evaluation Data/Motivation & \cellcolor{niceorange!6.25!white}3 (6.2\%) & \cellcolor{niceorange!14.583333333333334!white}7 (14.6\%) & \cellcolor{softc0!90.00!white}0.90 & \cellcolor{softc0!89.00!white}0.89 \\ 
		Evaluation Data/Preprocessing & \cellcolor{niceorange!4.166666666666666!white}2 (4.2\%) & \cellcolor{niceorange!4.166666666666666!white}2 (4.2\%) & \cellcolor{softc0!97.00!white}0.97 & \cellcolor{softc0!100.00!white}1.00 \\ 
		Training Data/Datasets & \cellcolor{niceorange!0.0!white}0 (0.0\%) & \cellcolor{niceorange!10.416666666666668!white}5 (10.4\%) & \cellcolor{softc0!95.00!white}0.95 & \cellcolor{softc0!98.00!white}0.98 \\ 
		Training Data/Motivation & \cellcolor{niceorange!2.083333333333333!white}1 (2.1\%) & \cellcolor{niceorange!16.666666666666664!white}8 (16.7\%) & \cellcolor{softc0!86.00!white}0.86 & \cellcolor{softc0!85.00!white}0.85 \\ 
		Training Data/Preprocessing & \cellcolor{niceorange!0.0!white}0 (0.0\%) & \cellcolor{niceorange!8.333333333333332!white}4 (8.3\%) & \cellcolor{softc0!96.00!white}0.96 & \cellcolor{softc0!98.00!white}0.98 \\ 
		Quantitative Analyses/Unitary results & \cellcolor{niceorange!8.333333333333332!white}4 (8.3\%) & \cellcolor{niceorange!4.166666666666666!white}2 (4.2\%) & \cellcolor{softc0!93.00!white}0.93 & \cellcolor{softc0!95.00!white}0.95 \\ 
		Quantitative Analyses/Intersectional results & \cellcolor{niceorange!0.0!white}0 (0.0\%) & \cellcolor{niceorange!4.166666666666666!white}2 (4.2\%) & \cellcolor{softc0!98.00!white}0.98 & \cellcolor{softc0!100.00!white}1.00 \\ 
		Memory or Hardware Requirements/Loading Requirements & \cellcolor{niceorange!2.083333333333333!white}1 (2.1\%) & \cellcolor{niceorange!4.166666666666666!white}2 (4.2\%) & \cellcolor{softc0!93.00!white}0.93 & \cellcolor{softc0!93.00!white}0.93 \\ 
		Memory or Hardware Requirements/Deploying Requirements & \cellcolor{niceorange!0.0!white}0 (0.0\%) & \cellcolor{niceorange!6.25!white}3 (6.2\%) & \cellcolor{softc0!93.00!white}0.93 & \cellcolor{softc0!95.00!white}0.95 \\ 
		Memory or Hardware Requirements/Training or Fine-tuning Requirements & \cellcolor{niceorange!0.0!white}0 (0.0\%) & \cellcolor{niceorange!8.333333333333332!white}4 (8.3\%) & \cellcolor{softc0!96.00!white}0.96 & \cellcolor{softc0!97.00!white}0.97 \\ 
		Ethical Considerations & \cellcolor{niceorange!2.083333333333333!white}1 (2.1\%) & \cellcolor{niceorange!31.25!white}15 (31.2\%) & \cellcolor{softc0!94.00!white}0.94 & \cellcolor{softc0!95.00!white}0.95 \\ 
		Caveats and Recommendations/Caveats & \cellcolor{niceorange!0.0!white}0 (0.0\%) & \cellcolor{niceorange!14.583333333333334!white}7 (14.6\%) & \cellcolor{softc0!88.00!white}0.88 & \cellcolor{softc0!88.00!white}0.88 \\ 
		Caveats and Recommendations/Recommendations & \cellcolor{niceorange!4.166666666666666!white}2 (4.2\%) & \cellcolor{niceorange!10.416666666666668!white}5 (10.4\%) & \cellcolor{softc0!89.00!white}0.89 & \cellcolor{softc0!89.00!white}0.89 \\ 
        \midrule
        \multirow{2}{*}{\textbf{Across all (sub)sections}} & \multicolumn{1}{c}{\textbf{Total:}} & \multicolumn{1}{c}{\textbf{Total:}} & \textbf{Avg: 0.94} & \textbf{Avg: 0.95} \\
         & \multicolumn{1}{c}{\textbf{28 (1.9\%)}} & \multicolumn{1}{c}{\textbf{147 (10.2\%)}} & \textbf{Med: 0.94} & \textbf{Med: 0.95} \\
    \bottomrule
\end{tabular}

%% file: result-reorganization-consistency.tex
\subsection{RQ1.3: \reorganizationrqconsistent}
\label{sec:rq_consistency}

\textit{Motivation:}
As LLMs are inherently non-deterministic, the contents in the model cards can be reorganized differently in each run. Therefore, we examined the consistency of content reorganization in model cards to assess whether \reorganizer{}'s performance remains stable across runs or exhibits notable variation. This analysis provides insight into the impact of stochasticity on content reorganization.

\textit{Approach:}
We evaluated the consistency of model card reorganization by measuring semantic similarity across multiple runs. We ran \reorganizer three times using the same settings and compared the resulting model cards (1) as complete documents and (2) at the level of individual sections. We prioritize semantic similarity over lexical similarity, as our goal is to assess whether contents convey the same meaning rather than use similar wording.  

To measure semantic similarity, we used cosine similarity between sentence embeddings generated by the \texttt{\seqsplit{allenai/specter}}~\cite{cohan2020specter} model, which is fine-tuned for capturing semantic similarity in scientific text. For each model card, we calculated cosine similarity scores between each pair of reorganized versions: run 1 vs. run 2, run 2 vs. run 3, and run 1 vs. run 3. We then averaged these three scores to obtain a single semantic similarity score per model card (hence 48 scores in total). From this distribution, we computed the overall average and median semantic similarity across all model cards, which reflects the general consistency of the reorganization process in preserving the semantic content across multiple runs. 
Cosine similarity scores near 1 represent identical or highly similar sentences across runs, and scores near 0 represent no semantic similarity~\cite{abbas2023relationship}. 

\begin{figure}[tb]
    \centering
    \includegraphics[width=.6\columnwidth]{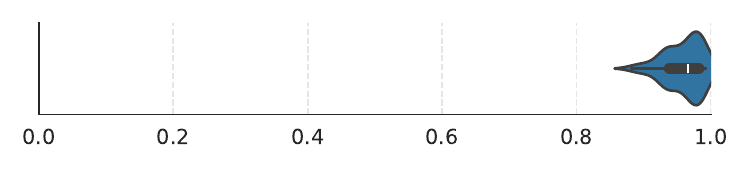}
    \caption{Distribution of document level semantic similarity scores across 48 model cards}
    \label{fig:document_level_semantic_similarity_scores_distribution}
\end{figure}

\textit{Findings:}
\textbf{We observed a high degree of consistency in the reorganized model cards across multiple runs}. From~\autoref{fig:document_level_semantic_similarity_scores_distribution}, we see that the median of the average semantic similarity scores across the three runs is 0.97, indicating that \reorganizer reliably preserved the overall meaning of the model cards across runs. 

\textbf{Our section-level analysis revealed that \reorganizer produced high semantic consistency across multiple runs}. From~\autoref{tab:merged_table}, we see that most (87.5\%) (sub)sections achieved semantic similarity scores $\ge$ 0.90. In particular, structured or well-scoped sections such as \texttt{Citation details}, \texttt{License}, \texttt{Contact}, \texttt{How to Use}, and \texttt{Intersectional results} exhibited near-perfect semantic similarity ($\ge$ 0.98 median, $\ge$ 0.97 average), indicating that \reorganizer consistently preserved the intended meaning of content in these sections. These sections are very straightforward and some are typically short, which likely contributes to their high stability. The  \texttt{Decision thresholds} and \texttt{Evaluation Data/Preprocessing} sections also show near-perfect semantic similarity, but these are mostly empty (``Not available.'').




\begin{tcolorbox}
The reorganization process demonstrated high consistency, with a median of the average semantic similarity scores across three runs is 0.97 for the full documents. Most (87.5\%) (sub)sections achieved semantic similarity scores of 0.90 or higher, indicating that \reorganizer reliably preserves meaning across runs. This high level of consistency suggests that reorganization by \reorganizer is not a product of random generation but rather reflects stable and intentional output patterns, strengthening confidence in the approach's reliability for practical model card reorganization task.
\end{tcolorbox}

%% file: discussion-reorganization.tex
\subsection{Discussion}
\label{sec:discussion}
\reorganizer is a highly viable solution for real-world use in standardizing model card documentation with only limited human oversight. The accuracy of content placement into the correct sections is high, with few hallucinations and misplacements. Also, the reorganizations produced by \reorganizer remain highly consistent across multiple runs. These findings collectively indicate \reorganizer's stable and reliable behavior, making it a dependable tool for automated model card reorganization.

Across the 48 reorganized model cards, manual effort was required in two areas: (1)~completing missing information based on original content, and (2)~correcting hallucinated or misinterpreted content. For completion, a median of only 4.5\% of checklist items had to be manually added, with a median normalized Levenshtein distance of 0.03 between the reorganized and completed versions, indicating light textual addition. While adding some of the missing information, such as usage guidance, explanatory or subjective content, contributes to user understanding and effective model use, some was not always strictly necessary, such as introductory blurbs and navigation phrases. 

In terms of hallucinations, only 147 out of 1,440 (sub)sections (10.2\%) needed manual edits. Most of these hallucinations occurred in descriptive or subjective sections, such as \texttt{Primary Intended Users} and \texttt{Ethical Considerations}, where the model occasionally inferred roles or intentions not explicitly stated in the original model card. These additions were typically vague and speculative, rather than technical, and rarely introduced factual inaccuracies. As a result, they posed minimal risk to the integrity of the documentation and were generally easy to identify and correct.


During our manual analysis of the reorganized model cards, we observed that content placed under the ``Additional Information'' section was often more appropriately aligned with existing standard sections. In particular, images were frequently relegated to this catch-all section rather than being contextually integrated into the relevant parts of the model card. These findings suggest that \reorganizer adopts a conservative strategy when encountering uncertain or partially understood content. While the ``Additional Information'' section helps avoid incorrect placements, it may reduce the clarity and utility of the reorganized model card. Future improvements could involve systematically further prompting \reorganizer to reassess and reassign such content to the most appropriate predefined sections within the model card structure. Recent works~\cite{zheng2023progressive, krishna2024understanding, chang2024efficient} support the idea that iterative prompting (asking the LLM to reconsider or refine its initial answer) consistently yields better results than single‑shot prompting.




%% file: result-generation.tex
\section{Evaluation of \generator}
\label{sec:result_generation}
In this section, we present the evaluation of our LLM-based model card generation approach, \generator.

\input{result-generation-resemble-original}
\input{result-generation-correct}
\input{result-generation-vary-input}

%% file: result-generation-resemble-original.tex
\subsection{RQ2.1: \generationrqresemble}
\label{sec:generationrqresemble}

\textit{Motivation:}
This will help us understand how close the information in the generated model cards is to the information in the original model cards, to understand how much \generator can accurately reproduce or infer key details when generating model cards from available repository data.

\textit{Approach:}
We evaluated the closeness of the generated model cards to the corresponding original model cards by measuring the semantic similarity between the generated and reorganized model cards. Like in Section~\ref{sec:rq_consistency}, we compared the model cards (1) as complete documents and (2) at the level of individual sections. We used cosine similarity between sentence embeddings of the model cards generated by the \texttt{allenai/specter}~\cite{cohan2020specter} model to measure semantic similarity. For each model, we calculated the cosine similarity between the generated and reorganized model cards, and then computed the overall average and median similarity scores across all models to reflect their general semantic closeness. Cosine similarity scores close to 1 indicate that the sentences in the model cards are highly similar or nearly identical, whereas scores close to 0 indicate little to no semantic similarity~\cite{abbas2023relationship}. We further examined the extent to which sections have empty or missing (marked as \texttt{Insufficient information} during generation) in the generated model cards, irrespective of having content in the original model card.

We did not compare the original model cards with their reorganized counterparts using semantic similarity because the reorganization step is structure-preserving rather than content-generating. The goal of reorganization is to redistribute existing information across a standardized template without introducing, removing, or modifying content. As such, the original and reorganized model cards are expected to be semantically equivalent by content, differing only in layout and section placement. Consequently, a semantic similarity evaluation between these two versions would be uninformative and potentially misleading, as any measured differences would primarily reflect structural changes rather than content discrepancies. In contrast, comparing the original model cards with the model cards generated from repository data directly evaluates the model’s ability to recover and synthesize model card content, which is the primary objective of our evaluation.

\begin{figure}[tb]
    \centering
    \includegraphics[width=.6\linewidth]{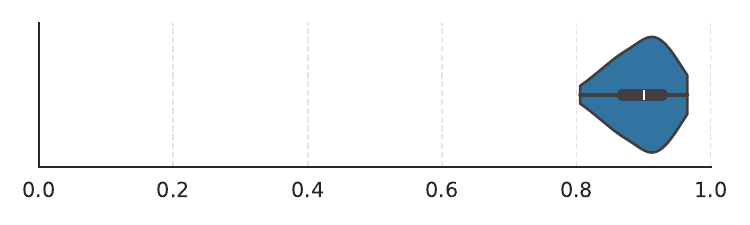}
    \caption{Distribution of semantic similarity scores of the 48 generated model cards with corresponding original (reorganized) model cards}
    \label{fig:distribution_of_model_card_similarity_scores}
\end{figure}

\textit{Findings:}
On average, \textbf{the generated model cards exhibited a high degree of semantic similarity to the original model cards}. From~\autoref{fig:distribution_of_model_card_similarity_scores}, we can see that when the generated model cards are compared as complete documents with the corresponding original model cards, the mean cosine similarity was 0.9, with a median of 0.9, indicating that the generated model cards generally preserved the key informational content of the originals.

\textbf{At the section level, similarity varied across sections}. Across all (sub)sections, the cosine similarity scores ranged from 0.62 (minimum similarity across all (sub)sections) to 0.99 (maximum similarity score across all (sub)sections), and the average of the average similarity scores for each (sub)section across the 48 model cards was 0.84. Empty sections or sections with limited content, such as \texttt{Decision thresholds}, \texttt{Intersectional results}, \texttt{Citation details}, and \texttt{Variation approaches}, achieved the highest similarity scores (averaging 0.88–0.91), likely because both the original and generated model cards contained minimal or no substantive information in these sections. In contrast, sections requiring interpretation, synthesis, or contextual reasoning, such as \texttt{Recommendations}, \texttt{Motivation} (under both \texttt{Evaluation Data} and \texttt{Training Data}), and \textit{Caveats}, showed comparatively lower similarity scores (averaging 0.76–0.80). This indicates that, while these sections are inherently more challenging to replicate fully, the model cards still capture a significant portion of the content from the original model card. Importantly, this highlights both the practical utility of automated generation and the opportunity to further improve coverage in these nuanced sections through targeted guidance, structured templates, or supplementary prompts.

When we analyzed the empty sections in the generated model cards, from~\autoref{tab:empty_section_comparison_table}, we observed that, compared to the reorganized model cards, the number of empty sections in the generated model cards increased for 16 (sub)sections and decreased for 13 (sub)sections. In all cases, the changes are less than 30 percentage points, except for the \texttt{Evaluation Data/Preprocessing} subsection, where the proportion of empty subsections decreased by 33.3 percentage points.

\begin{table}[]
    \centering
    \small
    \caption{(Sub)section-wise percentage of empty sections across 48 model cards. Numbers inside ``()'' indicate the total number of empty counts for the corresponding (sub)section among the 48 model cards. Darker orange colors indicate worse outcomes (a higher number of empty sections generated compared to the reorganized model cards), while darker blue colors indicate better outcomes (a lower number of empty sections generated compared to the reorganized model cards).}
    \label{tab:empty_section_comparison_table}
    \input{fig/empty_section_comparison_table}
\end{table}

\begin{tcolorbox}
The generated model cards showed a high degree of semantic similarity to the original ones, with a median cosine similarity score of 0.9. However, section-level analysis revealed a pattern: sections containing limited or straightforward data tended to have higher similarity, whereas sections requiring interpretation, synthesis, or contextual reasoning exhibited only moderate to high similarity.
\end{tcolorbox}

%% file: fig/empty_section_comparison_table.tex
\begin{tabular}{@{}lrr@{}}
	\toprule
	\textbf{(Sub)sections} & \textbf{Reorganized} & \textbf{Generated} \\
	\midrule
		Model Details/Person or organization developing model & 2.1\% ( 1) & \cellcolor{niceorange!20.833333333333336!white}$+$ 10.4\% ( 6) \\
		Model Details/Model date & 37.5\% (18) & \cellcolor{softc0!16.666666666666664!white}$-$ 8.3\% (14) \\
		Model Details/Model version & 10.4\% ( 5) & \cellcolor{softc0!8.333333333333336!white}$-$ 4.2\% ( 3) \\
		Model Details/Model type & 0.0\% ( 0) & 0.0\% ( 0) \\
		Model Details/Training details & 4.2\% ( 2) & 0.0\% ( 2) \\
		Model Details/Paper or other resource for more information & 4.2\% ( 2) & \cellcolor{niceorange!54.16666666666667!white}$+$ 27.1\% (15) \\
		Model Details/Citation details & 29.2\% (14) & \cellcolor{niceorange!58.333333333333336!white}$+$ 29.2\% (28) \\
		Model Details/License & 29.2\% (14) & \cellcolor{niceorange!45.833333333333336!white}$+$ 22.9\% (25) \\
		Model Details/Contact & 56.2\% (27) & \cellcolor{softc0!25.0!white}$-$ 12.5\% (21) \\
		Intended Use/Primary intended uses & 0.0\% ( 0) & 0.0\% ( 0) \\
		Intended Use/Primary intended users & 29.2\% (14) & \cellcolor{softc0!4.166666666666671!white}$-$ 2.1\% (13) \\
		Intended Use/Out-of-scope uses & 39.6\% (19) & \cellcolor{softc0!12.5!white}$-$ 6.2\% (16) \\
		How to Use & 0.0\% ( 0) & \cellcolor{niceorange!37.5!white}$+$ 18.8\% ( 9) \\
		Factors/Relevant factors & 31.2\% (15) & \cellcolor{niceorange!8.333333333333343!white}$+$ 4.2\% (17) \\
		Factors/Evaluation factors & 56.2\% (27) & \cellcolor{softc0!33.33333333333334!white}$-$ 16.7\% (19) \\
		Metrics/Model performance measures & 14.6\% ( 7) & \cellcolor{niceorange!33.33333333333333!white}$+$ 16.7\% (15) \\
		Metrics/Decision thresholds & 93.8\% (45) & \cellcolor{softc0!50.0!white}$-$ 25.0\% (33) \\
		Metrics/Variation approaches & 70.8\% (34) & \cellcolor{softc0!45.83333333333334!white}$-$ 22.9\% (23) \\
		Evaluation Data/Datasets & 22.9\% (11) & \cellcolor{niceorange!25.000000000000014!white}$+$ 12.5\% (17) \\
		Evaluation Data/Motivation & 60.4\% (29) & \cellcolor{softc0!33.33333333333333!white}$-$ 16.7\% (21) \\
		Evaluation Data/Preprocessing & 75.0\% (36) & \cellcolor{softc0!66.66666666666666!white}$-$ 33.3\% (20) \\
		Training Data/Datasets & 6.2\% ( 3) & \cellcolor{niceorange!58.33333333333334!white}$+$ 29.2\% (17) \\
		Training Data/Motivation & 35.4\% (17) & \cellcolor{niceorange!16.666666666666657!white}$+$ 8.3\% (21) \\
		Training Data/Preprocessing & 41.7\% (20) & \cellcolor{softc0!54.16666666666667!white}$-$ 27.1\% ( 7) \\
		Quantitative Analyses/Unitary results & 29.2\% (14) & \cellcolor{niceorange!25.000000000000007!white}$+$ 12.5\% (20) \\
		Quantitative Analyses/Intersectional results & 91.7\% (44) & \cellcolor{softc0!12.5!white}$-$ 6.2\% (41) \\
		Memory or Hardware Requirements/Loading Requirements & 62.5\% (30) & \cellcolor{softc0!50.0!white}$-$ 25.0\% (18) \\
		Memory or Hardware Requirements/Deploying Requirements & 58.3\% (28) & \cellcolor{niceorange!12.500000000000014!white}$+$ 6.3\% (31) \\
		Memory or Hardware Requirements/Training or Fine-tuning Requirements & 35.4\% (17) & \cellcolor{niceorange!12.5!white}$+$ 6.2\% (20) \\
		Ethical Considerations & 27.1\% (13) & \cellcolor{niceorange!8.333333333333336!white}$+$ 4.2\% (15) \\
		Caveats and Recommendations/Caveats & 6.2\% ( 3) & \cellcolor{niceorange!37.5!white}$+$ 18.8\% (12) \\
		Caveats and Recommendations/Recommendations & 6.2\% ( 3) & \cellcolor{niceorange!33.33333333333333!white}$+$ 16.7\% (11) \\
	\bottomrule
\end{tabular}

%% file: result-generation-correct.tex
\subsection{RQ2.2: \generationrqcorrect}

\textit{Motivation:}
While in RQ2.1, we evaluated semantic similarity between the original and generated model cards to assess overall content preservation, high similarity does not guarantee factual correctness. Large language models can produce text that is semantically coherent and closely aligned with the source while still introducing unsupported, speculative, or fabricated details. Therefore, in this research question, we focus specifically on the factual correctness of the generated model cards. We examine the extent to which information in the generated cards is explicitly supported by the original repository files, helping to assess the factual reliability of \generator.

\textit{Approach:}
For each generated model card, we used an LLM jury to evaluate whether the information in each (sub)section of the model card was factually correct based on the corresponding model repository files. We used \texttt{Gemini 2.5 Pro} and \texttt{Claude Sonnet 4.5} as the two primary jurors because of their extended context window capabilities (approximately 2 million and 1 million tokens, respectively). In cases of disagreement between these two jurors, we used \texttt{GPT-5} (supporting approximately 400,000 tokens) as a third juror to resolve the conflict. All three jurors are listed within the top 10 positions on the Text Arena leaderboard\footnote{\url{https://lmarena.ai/leaderboard/text}}. The context window size matters for this evaluation, as the models need to search all provided model resources to verify whether generated information was correct. 

Initially, we provided the jurors with all model repository files for evaluation. However, some requests exceeded the token limit, resulting in errors. To address this, we instead supplied the jurors only with the specific repository files cited in the generated model cards and asked them to evaluate each section's factual accuracy based on those files. 

When a juror marked a section as incorrect, we also asked it to specify what information was incorrect. However, as in Section~\ref{sec:reorganization_rq_correct}, we did not manually verify the information labeled as incorrect. Confirming whether these instances were truly incorrect would require detailed domain expertise and extensive manual inspection of numerous model repository files. Whenever two jurors agreed that a (sub)section contained incorrect information, we considered that (sub)section incorrect.


We counted the total number of model cards generated without any incorrect (sub)sections in them to understand the capability of \generator to produce model cards without any factually incorrect information. We also manually analyzed the incorrect information in the sections to understand the types of inaccuracies in information generation.

\begin{figure}[tb]
    \centering
    \includegraphics[width=.6\columnwidth]{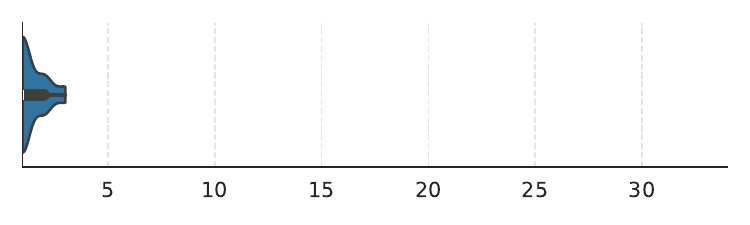}
    \caption{Distribution of the number of incorrect (sub)sections per model card among the 21 incorrect model cards. Note: One model card has 34 (sub)sections.}
    \label{fig:distribution_of_no_of_incorrect_sections}
\end{figure}

\textit{Findings:}
\textbf{More than half of the generated model cards were entirely correct, while the remaining ones contained only a small number of factual inaccuracies.} Twenty-six (54.17\%) model cards had all correct sections based on the jury result. From~\autoref{fig:distribution_of_no_of_incorrect_sections}, we can see that the average number of incorrect sections in the remaining 22 model cards is 1.41 (4.14\% of the total number of sections in a model card), with a median of 1.0 and a range from 1 to 3.

A manual analysis of the incorrect information across the generated model cards identified by the jury revealed several recurring patterns in the types and sources of factual inaccuracies. \textbf{Most incorrect information (60\%) arises from instances where the content in the model card does not accurately reflect the information in the cited sources}. These cases typically involve numerical or factual discrepancies between what the model card states and what is found in the associated repository files. Common examples include incorrect dataset sizes (e.g., \texttt{BAAI/bge-m3}, \texttt{FacebookAI/roberta-large-mnli}), misreported hyperparameters (e.g., \texttt{mosaicml/mpt-7b}, \texttt{\seqsplit{cerebras/Cerebras-GPT-13B}}), or conflicting learning rate schedules. Interestingly, some mismatches occur due to information mismatches in the sources themselves. For example, in \texttt{openai/whisper-large-v3}, the generated model card states the encoder processes an 80-channel log-Mel spectrogram, citing the associated paper. However, the \texttt{config.json} file specifies \texttt{num\_mel\_bins: 128}, indicating it uses a 128-channel spectrogram. This type of mismatch suggests that information consistency in the repository files is a key factor for the factual correctness of the generated model cards.

\textbf{The second most common inaccuracies result from hallucination (about 23\%)}, where \generator makes interpretive or inferential claims not directly supported by the source material. For example, \texttt{beomi/llama-2-ko-7b} attributes language specialization to the model name without evidence, and \texttt{Tencent-Hunyuan/HunyuanDiT} and \texttt{DeepFloyd/IF-I-XL-v1.0} speculate about dataset use or evaluation metrics without textual proof. This suggests that \generator tends to “fill in” missing details with plausible but unverifiable statements, a behavior consistent with known LLM tendencies toward hallucination when contextual grounding is weak.

The next pattern, accounting for about 17\% of the total, involves cases where the jury indicated that a model card references a document or file that does not exist or cannot be verified. In two instances (8.5\%) (\texttt{jbetker/tortoise-tts-v2} and \texttt{facebook/seamless-m4t-v2-large}), this issue occurred because we had to remove or truncate certain files before providing them to the juror models, as not all file types were supported across all LLMs, resulting in apparent citation mismatches. In two other cases (8.5\%), the cited information was drawn from multiple sources that contained conflicting details, making the citation valid for one source but invalid for another. For example, \texttt{bigscience/bloom} lists the vocabulary size as 250,880 and cites both \texttt{config.json} and the associated paper. While \texttt{config.json} confirms the vocabulary size, the paper reports it as 250,680. As a result, citing the paper to support the 250,880 value is incorrect. This finding illustrates that even when the factual content is technically correct, weak or inconsistent citation grounding can compromise transparency and reproducibility, both of which are core objectives of model documentation.

\begin{tcolorbox}
 More than half of the generated model cards (54.17\%) were fully factually correct. Among the remaining cards, the amount of incorrect information was minimal, as the median model card contained only one (sub)section with an inaccuracy. A manual analysis of these cases showed that, while \generator effectively maintains a coherent structure and preserves citations, it sometimes struggles with factual alignment between the generated content and the referenced files. In several cases, this misalignment was caused due to conflicting information in the referenced files.
\end{tcolorbox}

%% file: result-generation-vary-input.tex
\subsection{RQ2.3: \generationrqvaryinginput}
\label{sec:generationrqvaryinginput}

\textit{Motivation:}
Identifying the most important input data allows users to focus on the relevant repository files when using \generator. By providing only the necessary data, users can reduce monetary costs and stay within the model’s input token limits, while still ensuring that the generated model card captures the essential information.

\textit{Approach:}
Model repositories contain various files stored in a version control system, and the type of information contained in these files can influence the quality of the information generated in model cards. To examine this impact, we conducted a series of ablation analyses. We first identified the files cited within the generated model cards, as these represent the files actually used by \generator during model card generation. We then categorized the cited files by information type and regenerated the model cards using \generator under the same settings, but with each information type removed in turn from the repository files. Finally, we compared the regenerated model cards with the originally generated ones and calculated the cosine similarity scores to measure how much they differed.

Our earlier consistency analysis (Section~\ref{sec:rq_consistency}) showed that reorganized versions of the same model card achieve a very high similarity score of 0.97. Therefore, in this context, we consider similarity scores between 0.97 and 1.0 to indicate that the regenerated model cards preserve the same semantic content as the original generated versions, without information loss.

We also performed a section-level comparison to assess the impact of each information type in more detail. For each (sub)section, we calculated the cosine similarity between the generated and regenerated model cards. In addition, we examined whether removing a particular information type resulted in any (sub)section being empty, which was labeled as ``Insufficient Information" during generation. We then analyzed these results across all repositories to identify patterns in how different information types influenced model card quality and to determine which types consistently contributed to more informative sections.

\textit{Findings:}
\textbf{Across the analyzed 48 repositories, configuration files (e.g., config.json), tokenizer files (e.g., tokenizer.json), and associated papers were the most commonly cited information sources.} Associated papers exhibited the highest citation rate: all 26 repositories that included a paper consistently cited it in the generated model cards, making it the most frequently referenced information type. Configuration files were present in nearly all repositories (46) and were cited in every case in which they appeared. Tokenizer files were cited less consistently; although 43 repositories contained a tokenizer file, only 34 (79.1\%) cited them in the generated model cards. Additional repositories contained tokenizer files stored using LFS due to their large size; these files were excluded from our analysis.

\begin{landscape}
\begin{table}[tb]
    \small
    \caption{(Sub)section-wise semantic similarity score comparison. Darker colors mean better outcomes (higher similarity)}
    \label{tab:data_variation_similarity_comparison_table}
    \centering
	\input{fig/data_variation_similarity_comparison_table}
\end{table}
\end{landscape}

\begin{landscape}
\begin{table}[tb]
    \small
    \caption{(Sub)section-wise comparison of empty sections expressed as percentages. The first value inside ``()'' demotes the total number of empty counts for the corresponding (sub)section, while the second value denotes the total number of model cards. The empty section counts for the originally generated model cards (`w All Data' column) are same to those reported in the `Generated' column of~\autoref{tab:empty_section_comparison_table} and are reported here for ease of comparison. Darker colors indicate worse outcomes (more empty sections without the data type in comparison with providing all repository data). For example, the prevalence of darker cells in the `w/o Paper' column indicate that removing the paper considerably increases the number of empty sections.}
    \label{tab:insufficient_information_comparison_table}
    \centering
	\input{fig/insufficient_information_comparison_table}
\end{table}
\end{landscape}

\textbf{Excluding associated papers during model card regeneration resulted in a noticeable loss of information compared to the originally generated model cards.} Regenerating the model cards without these papers resulted in a median document level cosine similarity of 0.91 when compared to the originally generated model cards. At the (sub)section level, as shown in~\autoref{tab:data_variation_similarity_comparison_table}, we observe that for a median regenerated model card without its corresponding paper, all (sub)sections, except five, have similarity scores below 0.91. From~\autoref{tab:insufficient_information_comparison_table}, we see that, for the sections \texttt{Model Details/Citation details}, \texttt{Training Data/Motivation} and \texttt{Quantitative Analyses/Intersectional results}, \generator was unable to generate any information in the absence of the paper, whereas, when all repository data were available, these sections contained insufficient information in 58.3\%, 43.8\% and 85.4\% of the cases, respectively. This suggests that the paper often serves as the primary and most comprehensive source of contextual details.

Next, we found that \textbf{excluding the configuration files during model card regeneration resulted in an information loss, though the impact was smaller than that observed when associated papers were omitted.} Regenerating the model cards without these configuration files resulted in a median document level cosine similarity of 0.95 when compared to the originally generated model cards. At the (sub)section level, as shown in~\autoref{tab:data_variation_similarity_comparison_table}, we observe that for regenerated model cards without configuration files, the median similarity scores range from 0.90 to 1.00. Except for 6 (sub)sections, all the median scores are $\ge$0.95. Furthermore, from~\autoref{tab:insufficient_information_comparison_table}, we see that the number of empty (sub)sections does not vary significantly when configuration files are removed. This is likely because the presence of the associated paper reduces reliance on configuration files as a primary information source.

We found that \textbf{regenerating the model cards without tokenizer files had a similar impact to regeneration without configuration files.} Excluding tokenizer files resulted in a median document level cosine similarity of 0.96, closely matching the similarity observed when configuration files were omitted. Consistent with this result, the (sub)section-level analysis revealed comparable patterns between model cards regenerated without tokenizer files and those regenerated without configuration files.


\begin{tcolorbox}
    The corresponding papers of the repositories have the greatest impact on the content generation of \generator. While configuration and tokenizer files are also frequently cited, their absence has relatively little effect on \generator{}'s output, likely because the presence of the associated paper already provides overlapping or more comprehensive information.
\end{tcolorbox}

%% file: fig/data_variation_similarity_comparison_table.tex
\begin{tabular}{@{}lrrrrrr@{}}
	\toprule
	\multirow{2}{*}{\textbf{(Sub)sections}} & \multicolumn{2}{c}{\textbf{w/o Paper}} & \multicolumn{2}{c}{\textbf{w/o Config}} & \multicolumn{2}{c}{\textbf{w/o Tokenizer}} \\
	\cmidrule(lr){2-3} \cmidrule(lr){4-5} \cmidrule(lr){6-7}
	 & \textbf{{Avg}} & \textbf{{Med}} & \textbf{{Avg}} & \textbf{{Med}} & \textbf{{Avg}} & \textbf{{Med}} \\
	\midrule
		Model Details/Person or organization developing model & \cellcolor{softc0!86.56!white}0.87 & \cellcolor{softc0!84.46!white}0.84 & \cellcolor{softc0!94.82!white}0.95 & \cellcolor{softc0!97.55!white}0.98 & \cellcolor{softc0!94.01!white}0.94 & \cellcolor{softc0!97.07!white}0.97 \\
		Model Details/Model date & \cellcolor{softc0!85.16!white}0.85 & \cellcolor{softc0!86.10!white}0.86 & \cellcolor{softc0!94.59!white}0.95 & \cellcolor{softc0!98.32!white}0.98 & \cellcolor{softc0!93.25!white}0.93 & \cellcolor{softc0!95.33!white}0.95 \\
		Model Details/Model version & \cellcolor{softc0!78.07!white}0.78 & \cellcolor{softc0!77.00!white}0.77 & \cellcolor{softc0!88.60!white}0.89 & \cellcolor{softc0!91.28!white}0.91 & \cellcolor{softc0!91.07!white}0.91 & \cellcolor{softc0!92.27!white}0.92 \\
		Model Details/Model type & \cellcolor{softc0!91.83!white}0.92 & \cellcolor{softc0!92.68!white}0.93 & \cellcolor{softc0!92.52!white}0.93 & \cellcolor{softc0!93.43!white}0.93 & \cellcolor{softc0!95.60!white}0.96 & \cellcolor{softc0!95.91!white}0.96 \\
		Model Details/Training details & \cellcolor{softc0!84.72!white}0.85 & \cellcolor{softc0!84.76!white}0.85 & \cellcolor{softc0!90.31!white}0.90 & \cellcolor{softc0!91.58!white}0.92 & \cellcolor{softc0!91.98!white}0.92 & \cellcolor{softc0!93.86!white}0.94 \\
		Model Details/Paper or other resource for more information & \cellcolor{softc0!75.29!white}0.75 & \cellcolor{softc0!75.43!white}0.75 & \cellcolor{softc0!95.53!white}0.96 & \cellcolor{softc0!97.73!white}0.98 & \cellcolor{softc0!95.32!white}0.95 & \cellcolor{softc0!97.29!white}0.97 \\
		Model Details/Citation details & \cellcolor{softc0!80.64!white}0.81 & \cellcolor{softc0!78.95!white}0.79 & \cellcolor{softc0!95.89!white}0.96 & \cellcolor{softc0!99.52!white}1.00 & \cellcolor{softc0!96.62!white}0.97 & \cellcolor{softc0!99.88!white}1.00 \\
		Model Details/License & \cellcolor{softc0!91.38!white}0.91 & \cellcolor{softc0!94.77!white}0.95 & \cellcolor{softc0!95.18!white}0.95 & \cellcolor{softc0!98.71!white}0.99 & \cellcolor{softc0!96.47!white}0.96 & \cellcolor{softc0!98.21!white}0.98 \\
		Model Details/Contact & \cellcolor{softc0!85.18!white}0.85 & \cellcolor{softc0!82.91!white}0.83 & \cellcolor{softc0!97.21!white}0.97 & \cellcolor{softc0!99.21!white}0.99 & \cellcolor{softc0!98.01!white}0.98 & \cellcolor{softc0!99.63!white}1.00 \\
		Intended Use/Primary intended uses & \cellcolor{softc0!88.06!white}0.88 & \cellcolor{softc0!89.15!white}0.89 & \cellcolor{softc0!93.00!white}0.93 & \cellcolor{softc0!94.79!white}0.95 & \cellcolor{softc0!94.35!white}0.94 & \cellcolor{softc0!95.29!white}0.95 \\
		Intended Use/Primary intended users & \cellcolor{softc0!78.96!white}0.79 & \cellcolor{softc0!78.80!white}0.79 & \cellcolor{softc0!92.10!white}0.92 & \cellcolor{softc0!94.64!white}0.95 & \cellcolor{softc0!91.59!white}0.92 & \cellcolor{softc0!93.27!white}0.93 \\
		Intended Use/Out-of-scope uses & \cellcolor{softc0!84.24!white}0.84 & \cellcolor{softc0!82.08!white}0.82 & \cellcolor{softc0!94.96!white}0.95 & \cellcolor{softc0!97.15!white}0.97 & \cellcolor{softc0!93.93!white}0.94 & \cellcolor{softc0!97.28!white}0.97 \\
		How to Use & \cellcolor{softc0!82.90!white}0.83 & \cellcolor{softc0!80.13!white}0.80 & \cellcolor{softc0!87.92!white}0.88 & \cellcolor{softc0!91.70!white}0.92 & \cellcolor{softc0!88.27!white}0.88 & \cellcolor{softc0!92.19!white}0.92 \\
		Factors/Relevant factors & \cellcolor{softc0!79.12!white}0.79 & \cellcolor{softc0!77.28!white}0.77 & \cellcolor{softc0!94.32!white}0.94 & \cellcolor{softc0!95.13!white}0.95 & \cellcolor{softc0!95.25!white}0.95 & \cellcolor{softc0!95.77!white}0.96 \\
		Factors/Evaluation factors & \cellcolor{softc0!77.53!white}0.78 & \cellcolor{softc0!78.02!white}0.78 & \cellcolor{softc0!95.30!white}0.95 & \cellcolor{softc0!96.85!white}0.97 & \cellcolor{softc0!93.51!white}0.94 & \cellcolor{softc0!95.82!white}0.96 \\
		Metrics/Model performance measures & \cellcolor{softc0!81.30!white}0.81 & \cellcolor{softc0!79.64!white}0.80 & \cellcolor{softc0!96.57!white}0.97 & \cellcolor{softc0!97.71!white}0.98 & \cellcolor{softc0!96.06!white}0.96 & \cellcolor{softc0!96.83!white}0.97 \\
		Metrics/Decision thresholds & \cellcolor{softc0!86.90!white}0.87 & \cellcolor{softc0!85.52!white}0.86 & \cellcolor{softc0!95.19!white}0.95 & \cellcolor{softc0!97.47!white}0.97 & \cellcolor{softc0!97.06!white}0.97 & \cellcolor{softc0!100.00!white}1.00 \\
		Metrics/Variation approaches & \cellcolor{softc0!81.70!white}0.82 & \cellcolor{softc0!82.78!white}0.83 & \cellcolor{softc0!95.24!white}0.95 & \cellcolor{softc0!97.21!white}0.97 & \cellcolor{softc0!94.44!white}0.94 & \cellcolor{softc0!95.58!white}0.96 \\
		Evaluation Data/Datasets & \cellcolor{softc0!78.62!white}0.79 & \cellcolor{softc0!78.36!white}0.78 & \cellcolor{softc0!96.67!white}0.97 & \cellcolor{softc0!98.04!white}0.98 & \cellcolor{softc0!97.34!white}0.97 & \cellcolor{softc0!98.44!white}0.98 \\
		Evaluation Data/Motivation & \cellcolor{softc0!73.91!white}0.74 & \cellcolor{softc0!74.32!white}0.74 & \cellcolor{softc0!94.56!white}0.95 & \cellcolor{softc0!95.68!white}0.96 & \cellcolor{softc0!95.13!white}0.95 & \cellcolor{softc0!95.68!white}0.96 \\
		Evaluation Data/Preprocessing & \cellcolor{softc0!79.19!white}0.79 & \cellcolor{softc0!77.70!white}0.78 & \cellcolor{softc0!92.80!white}0.93 & \cellcolor{softc0!95.27!white}0.95 & \cellcolor{softc0!94.75!white}0.95 & \cellcolor{softc0!97.59!white}0.98 \\
		Training Data/Datasets & \cellcolor{softc0!80.13!white}0.80 & \cellcolor{softc0!79.31!white}0.79 & \cellcolor{softc0!96.27!white}0.96 & \cellcolor{softc0!97.53!white}0.98 & \cellcolor{softc0!95.07!white}0.95 & \cellcolor{softc0!97.49!white}0.97 \\
		Training Data/Motivation & \cellcolor{softc0!72.71!white}0.73 & \cellcolor{softc0!72.45!white}0.72 & \cellcolor{softc0!94.26!white}0.94 & \cellcolor{softc0!95.58!white}0.96 & \cellcolor{softc0!95.78!white}0.96 & \cellcolor{softc0!98.49!white}0.98 \\
		Training Data/Preprocessing & \cellcolor{softc0!84.69!white}0.85 & \cellcolor{softc0!84.88!white}0.85 & \cellcolor{softc0!93.10!white}0.93 & \cellcolor{softc0!95.78!white}0.96 & \cellcolor{softc0!91.66!white}0.92 & \cellcolor{softc0!94.52!white}0.95 \\
		Quantitative Analyses/Unitary results & \cellcolor{softc0!74.50!white}0.74 & \cellcolor{softc0!72.91!white}0.73 & \cellcolor{softc0!94.86!white}0.95 & \cellcolor{softc0!96.77!white}0.97 & \cellcolor{softc0!94.69!white}0.95 & \cellcolor{softc0!97.25!white}0.97 \\
		Quantitative Analyses/Intersectional results & \cellcolor{softc0!88.81!white}0.89 & \cellcolor{softc0!91.73!white}0.92 & \cellcolor{softc0!94.97!white}0.95 & \cellcolor{softc0!97.94!white}0.98 & \cellcolor{softc0!94.12!white}0.94 & \cellcolor{softc0!100.00!white}1.00 \\
		Memory or Hardware Requirements/Loading Requirements & \cellcolor{softc0!88.78!white}0.89 & \cellcolor{softc0!90.86!white}0.91 & \cellcolor{softc0!91.63!white}0.92 & \cellcolor{softc0!94.58!white}0.95 & \cellcolor{softc0!92.47!white}0.92 & \cellcolor{softc0!95.54!white}0.96 \\
		Memory or Hardware Requirements/Deploying Requirements & \cellcolor{softc0!86.20!white}0.86 & \cellcolor{softc0!87.03!white}0.87 & \cellcolor{softc0!93.40!white}0.93 & \cellcolor{softc0!95.97!white}0.96 & \cellcolor{softc0!92.96!white}0.93 & \cellcolor{softc0!97.94!white}0.98 \\
		Memory or Hardware Requirements/Training or Fine-tuning Requirements & \cellcolor{softc0!84.99!white}0.85 & \cellcolor{softc0!85.38!white}0.85 & \cellcolor{softc0!94.88!white}0.95 & \cellcolor{softc0!96.87!white}0.97 & \cellcolor{softc0!93.81!white}0.94 & \cellcolor{softc0!96.01!white}0.96 \\
		Ethical Considerations & \cellcolor{softc0!86.38!white}0.86 & \cellcolor{softc0!86.26!white}0.86 & \cellcolor{softc0!94.67!white}0.95 & \cellcolor{softc0!96.16!white}0.96 & \cellcolor{softc0!95.58!white}0.96 & \cellcolor{softc0!96.58!white}0.97 \\
		Caveats and Recommendations/Caveats & \cellcolor{softc0!77.35!white}0.77 & \cellcolor{softc0!76.47!white}0.76 & \cellcolor{softc0!89.96!white}0.90 & \cellcolor{softc0!93.08!white}0.93 & \cellcolor{softc0!92.04!white}0.92 & \cellcolor{softc0!92.85!white}0.93 \\
		Caveats and Recommendations/Recommendations & \cellcolor{softc0!75.66!white}0.76 & \cellcolor{softc0!75.13!white}0.75 & \cellcolor{softc0!88.75!white}0.89 & \cellcolor{softc0!90.16!white}0.90 & \cellcolor{softc0!89.94!white}0.90 & \cellcolor{softc0!92.96!white}0.93 \\
    \midrule
        \multirow{2}{*}{\textbf{Across all (sub)sections}} & \textbf{Avg: 0.82} & \textbf{Avg: 0.82} & \textbf{Avg: 0.94} & \textbf{Avg: 0.96} & \textbf{Avg: 0.94} & \textbf{Avg: 0.96} \\
         & \textbf{Med: 0.82} & \textbf{Med: 0.81} & \textbf{Med: 0.95} & \textbf{Med: 0.96}  & \textbf{Med: 0.94} & \textbf{Med: 0.96} \\

	\bottomrule
\end{tabular}

%% file: fig/insufficient_information_comparison_table.tex
\begin{tabular}{@{}lrrrr@{}}
	\toprule
	\textbf{(Sub)sections} & \textbf{w All Data} & \textbf{w/o Paper} & \textbf{w/o Config} & \textbf{w/o Tokenizer} \\
	\midrule
		Model Details/Person or organization developing model & 12.5\% ( 6/48) & \cellcolor{niceorange!29.807692307692307!white}$+$ 29.8\% (11/26) & \cellcolor{niceorange!7.065217391304348!white}$+$ 7.1\% ( 9/46) & \cellcolor{niceorange!8.088235294117645!white}$+$ 8.1\% ( 7/34) \\
		Model Details/Model date & 29.2\% (14/48) & \cellcolor{niceorange!59.29487179487178!white}$+$ 59.3\% (23/26) & \cellcolor{niceorange!3.442028985507246!white}$+$ 3.4\% (15/46) & \cellcolor{niceorange!3.186274509803919!white}$+$ 3.2\% (11/34) \\
		Model Details/Model version & 6.2\% ( 3/48) & \cellcolor{niceorange!12.980769230769234!white}$+$ 13.0\% ( 5/26) & \cellcolor{niceorange!6.7934782608695645!white}$+$ 6.8\% ( 6/46) & $-$ 3.3\% ( 1/34) \\
		Model Details/Model type & 0.0\% ( 0/48) & 0.0\% ( 0/26) & 0.0\% ( 0/46) & 0.0\% ( 0/34) \\
		Model Details/Training details & 4.2\% ( 2/48) & $-$ 0.3\% ( 1/26) & \cellcolor{niceorange!21.92028985507246!white}$+$ 21.9\% (12/46) & \cellcolor{niceorange!1.715686274509804!white}$+$ 1.7\% ( 2/34) \\
		Model Details/Paper or other resource for more information & 31.2\% (15/48) & \cellcolor{niceorange!41.826923076923066!white}$+$ 41.8\% (19/26) & \cellcolor{niceorange!1.358695652173914!white}$+$ 1.4\% (15/46) & \cellcolor{niceorange!1.102941176470587!white}$+$ 1.1\% (11/34) \\
		Model Details/Citation details & 58.3\% (28/48) & \cellcolor{niceorange!41.666666666666664!white}$+$ 41.7\% (26/26) & \cellcolor{niceorange!2.536231884057976!white}$+$ 2.5\% (28/46) & $-$ 2.5\% (19/34) \\
		Model Details/License & 52.1\% (25/48) & \cellcolor{niceorange!20.99358974358973!white}$+$ 21.0\% (19/26) & \cellcolor{niceorange!4.438405797101446!white}$+$ 4.4\% (26/46) & \cellcolor{niceorange!9.681372549019606!white}$+$ 9.7\% (21/34) \\
		Model Details/Contact & 43.8\% (21/48) & \cellcolor{niceorange!44.71153846153845!white}$+$ 44.7\% (23/26) & $-$ 2.4\% (19/46) & \cellcolor{niceorange!6.25!white}$+$ 6.2\% (17/34) \\
		Intended Use/Primary intended uses & 0.0\% ( 0/48) & 0.0\% ( 0/26) & \cellcolor{niceorange!8.695652173913043!white}$+$ 8.7\% ( 4/46) & 0.0\% ( 0/34) \\
		Intended Use/Primary intended users & 27.1\% (13/48) & \cellcolor{niceorange!49.839743589743605!white}$+$ 49.8\% (20/26) & \cellcolor{niceorange!5.525362318840582!white}$+$ 5.5\% (15/46) & \cellcolor{niceorange!11.151960784313726!white}$+$ 11.2\% (13/34) \\
		Intended Use/Out-of-scope uses & 33.3\% (16/48) & \cellcolor{niceorange!39.74358974358974!white}$+$ 39.7\% (19/26) & \cellcolor{niceorange!10.14492753623189!white}$+$ 10.1\% (20/46) & \cellcolor{niceorange!13.725490196078432!white}$+$ 13.7\% (16/34) \\
		How to Use & 18.8\% ( 9/48) & \cellcolor{niceorange!23.557692307692307!white}$+$ 23.6\% (11/26) & \cellcolor{niceorange!2.9891304347826093!white}$+$ 3.0\% (10/46) & \cellcolor{niceorange!4.7794117647058805!white}$+$ 4.8\% ( 8/34) \\
		Factors/Relevant factors & 35.4\% (17/48) & \cellcolor{niceorange!45.3525641025641!white}$+$ 45.4\% (21/26) & \cellcolor{niceorange!5.887681159420282!white}$+$ 5.9\% (19/46) & \cellcolor{niceorange!8.700980392156858!white}$+$ 8.7\% (15/34) \\
		Factors/Evaluation factors & 39.6\% (19/48) & \cellcolor{niceorange!48.878205128205124!white}$+$ 48.9\% (23/26) & \cellcolor{niceorange!1.7210144927536248!white}$+$ 1.7\% (19/46) & \cellcolor{niceorange!7.475490196078432!white}$+$ 7.5\% (16/34) \\
		Metrics/Model performance measures & 31.2\% (15/48) & \cellcolor{niceorange!49.519230769230774!white}$+$ 49.5\% (21/26) & $-$ 0.8\% (14/46) & \cellcolor{niceorange!4.044117647058826!white}$+$ 4.0\% (12/34) \\
		Metrics/Decision thresholds & 68.8\% (33/48) & \cellcolor{niceorange!23.557692307692307!white}$+$ 23.6\% (24/26) & \cellcolor{niceorange!2.9891304347826093!white}$+$ 3.0\% (33/46) & \cellcolor{niceorange!10.661764705882348!white}$+$ 10.7\% (27/34) \\
		Metrics/Variation approaches & 47.9\% (23/48) & \cellcolor{niceorange!48.23717948717949!white}$+$ 48.2\% (25/26) & $-$ 0.1\% (22/46) & $-$ 0.9\% (16/34) \\
		Evaluation Data/Datasets & 35.4\% (17/48) & \cellcolor{niceorange!53.04487179487178!white}$+$ 53.0\% (23/26) & $-$ 0.6\% (16/46) & \cellcolor{niceorange!8.700980392156858!white}$+$ 8.7\% (15/34) \\
		Evaluation Data/Motivation & 43.8\% (21/48) & \cellcolor{niceorange!52.40384615384616!white}$+$ 52.4\% (25/26) & $-$ 0.3\% (20/46) & \cellcolor{niceorange!6.25!white}$+$ 6.2\% (17/34) \\
		Evaluation Data/Preprocessing & 41.7\% (20/48) & \cellcolor{niceorange!46.79487179487178!white}$+$ 46.8\% (23/26) & \cellcolor{niceorange!1.8115942028985472!white}$+$ 1.8\% (20/46) & \cellcolor{niceorange!5.3921568627450895!white}$+$ 5.4\% (16/34) \\
		Training Data/Datasets & 35.4\% (17/48) & \cellcolor{niceorange!29.967948717948715!white}$+$ 30.0\% (17/26) & \cellcolor{niceorange!3.713768115942024!white}$+$ 3.7\% (18/46) & \cellcolor{niceorange!5.7598039215686185!white}$+$ 5.8\% (14/34) \\
		Training Data/Motivation & 43.8\% (21/48) & \cellcolor{niceorange!56.25!white}$+$ 56.2\% (26/26) & $-$ 2.4\% (19/46) & \cellcolor{niceorange!6.25!white}$+$ 6.2\% (17/34) \\
		Training Data/Preprocessing & 14.6\% ( 7/48) & \cellcolor{niceorange!0.8012820512820511!white}$+$ 0.8\% ( 4/26) & \cellcolor{niceorange!2.807971014492752!white}$+$ 2.8\% ( 8/46) & \cellcolor{niceorange!20.71078431372549!white}$+$ 20.7\% (12/34) \\
		Quantitative Analyses/Unitary results & 41.7\% (20/48) & \cellcolor{niceorange!46.79487179487178!white}$+$ 46.8\% (23/26) & $-$ 4.7\% (17/46) & \cellcolor{niceorange!2.4509803921568576!white}$+$ 2.5\% (15/34) \\
		Quantitative Analyses/Intersectional results & 85.4\% (41/48) & \cellcolor{niceorange!14.583333333333343!white}$+$ 14.6\% (26/26) & $-$ 2.8\% (38/46) & $-$ 3.1\% (28/34) \\
		Memory or Hardware Requirements/Loading Requirements & 37.5\% (18/48) & \cellcolor{niceorange!12.5!white}$+$ 12.5\% (13/26) & \cellcolor{niceorange!16.847826086956516!white}$+$ 16.8\% (25/46) & \cellcolor{niceorange!6.617647058823529!white}$+$ 6.6\% (15/34) \\
		Memory or Hardware Requirements/Deploying Requirements & 64.6\% (31/48) & \cellcolor{niceorange!27.724358974358964!white}$+$ 27.7\% (24/26) & \cellcolor{niceorange!2.8079710144927503!white}$+$ 2.8\% (31/46) & \cellcolor{niceorange!6.004901960784309!white}$+$ 6.0\% (24/34) \\
		Memory or Hardware Requirements/Training or Fine-tuning Requirements & 41.7\% (20/48) & \cellcolor{niceorange!39.1025641025641!white}$+$ 39.1\% (21/26) & $-$ 0.4\% (19/46) & $-$ 0.5\% (14/34) \\
		Ethical Considerations & 31.2\% (15/48) & \cellcolor{niceorange!37.980769230769226!white}$+$ 38.0\% (18/26) & \cellcolor{niceorange!3.532608695652172!white}$+$ 3.5\% (16/46) & \cellcolor{niceorange!9.92647058823529!white}$+$ 9.9\% (14/34) \\
		Caveats and Recommendations/Caveats & 25.0\% (12/48) & \cellcolor{niceorange!28.846153846153847!white}$+$ 28.8\% (14/26) & $-$ 7.6\% ( 8/46) & \cellcolor{niceorange!1.4705882352941195!white}$+$ 1.5\% ( 9/34) \\
		Caveats and Recommendations/Recommendations & 22.9\% (11/48) & \cellcolor{niceorange!54.00641025641027!white}$+$ 54.0\% (20/26) & $-$ 1.2\% (10/46) & \cellcolor{niceorange!9.436274509803923!white}$+$ 9.4\% (11/34) \\
	\bottomrule
\end{tabular}

%% file: threats_to_validity.tex
\section{Threats to Validity}
\label{sec:threats_to_validity}


\textit{Internal Validity:} The first two authors of this paper manually reviewed and corrected the reorganized model cards and their corresponding checklists. Although we aimed to reduce subjectivity by having the authors work independently, the manual process remains inherently subjective and may vary depending on how each individual interprets the information. 

An internal threat to validity arises from inconsistencies and ambiguities in the repository files used to generate the model cards. Information about model characteristics is often scattered across multiple sources, which may be outdated or mutually inconsistent. As a result, the generated model cards may contain conflicting or incomplete information, potentially affecting their accuracy.

\textit{Construct Validity:} We assumed that model cards from top repositories on HF represent a high-quality baseline. Liang et al.~\cite{liang2024s} showed that the top 0.3\% (100 out of 32,111) repositories in HF have good model cards. In our study, to keep our dataset manageable for manual verification, we picked the top 0.1\% (1,000 out of 944,178) models. However, the quality of documentation within these repositories may still vary, and relying on them as our evaluation reference may have introduced bias.

The checklist items used to assess completeness vary in granularity. While most are sentences focused on a single piece of information, some are longer and contain multiple pieces of information. To address this inconsistency, we permitted partial matches during manual evaluation. However, the interpretation of partial completeness remains subjective and may introduce variability in the scoring process.

Further, to reduce human effort and scale the evaluation, we employed a jury of LLMs to assess the correctness of content placement within each section (\reorganizer) and the correctness of the content generation (\generator). While LLM juries offer a scalable and cost-effective way to simulate multi-rater evaluation, relying on them introduces potential threats to validity. Since LLMs are prone to hallucination and exhibit inherently non-deterministic behavior, this can reduce the reliability and reproducibility of the evaluation. Moreover, LLMs can lack deep comprehension and may rely on superficial patterns or heuristics, leading to assessments that diverge from expert human judgment. In addition, model biases can affect decision quality, especially in technical or context-sensitive scenarios. Finally, because LLMs offer limited transparency in their reasoning, it is difficult to audit their decisions or resolve disagreements. To mitigate these risks, we incorporated human validation, however, the initial use of LLMs in the evaluation process remains a potential threat. 

Reorganized content is sometimes placed in the ``Additional Information'' section by \reorganizer when it is unsure where to place that content. While this fallback helps avoid losing information, it can result in too much content being concentrated in that section. This behavior poses a threat to construct validity, as it may distort the intended organization of the content.

Finally, our study applied a uniform evaluation approach across all models, regardless of their type or application domain. 
Prior work by Richards et al.~\cite{richards2021human} highlights that model documentation practices can vary based on domain and model type, with some domains requiring more detailed or specialized information. By using a single standard model card template, we may have underrepresented domain-specific documentation patterns, potentially overlooking nuances that are important in certain specialized contexts.



\textit{External Validity:} Our ground truth for evaluating \reorganizer and \generator was based on a dataset of 48 model cards, which is not statistically representative of the broader population. This limited and selective sample could introduce bias that restricts the generalizability of our findings. In particular, our evaluation provides a strong indication of \reorganizer{}'s capabilities, although their performance on less curated or lower-quality model cards, which are more common in the broader ecosystem, may vary. Also, the capability of \generator was evaluated based on the repository files available in these 48 repositories. However, in a broader ecosystem, the types of resources contained in repertoires may vary and could influence \generator{}'s performance differently. These variations pose a potential threat to the external validity of our results.

%% file: related_work.tex
\section{Related Work}
\label{sec:related_work}
We group the related work by purpose and present it in the following two subsections.

\subsection{Model Documentation Analysis and Improvement}

Many studies examine the content of existing model documentation to identify gaps and the scope of improvements in model documentation. Bhat et al.~\cite{bhat2023aspirations} and Liang et al.~\cite{liang2024s} showed that model cards do not often contain enough information about different sections of the standard template. Pepe et al.~\cite{pepe2024hugging} highlighted the need for better documentation of training datasets, biases, and licenses in pre-trained models to improve transparency and mitigate potential biases and legal issues. Oreamuno et al.~\cite{oreamuno2024state} found that many models and datasets in the \HF store lack comprehensive documentation, either failing to meet user needs or lacking enforcement. The study demonstrated inconsistencies in ethics and transparency-related documentation for ML models and datasets, indicating the need for improved practices to address ethical concerns, biases, and limitations. Additionally, they suggested adding categories for model versioning and attribution in documentation standards. Gao et al.~\cite{gao2024documenting} particularly investigated how ethical aspects are currently documented in model cards and provided guidelines for improvement. Through thematic analysis of 256 model cards, they identified six key themes, with the most common being model behavioral risks, intended use cases, and risk mitigation strategies. 

Several studies have proposed recommendations for improving model documentation based on interviews, surveys and experience. To enhance ML accountability, Zainyte et al.~\cite{zainyte2021challenges} suggested implementing causality (the effect of model components on the system), decision provenance, and computational tests. They also emphasized the need for detailed factual information in places such as documentation to improve transparency and, in turn, accountability. Nunes et al.~\cite{nunes2022using}, through a qualitative study with developers, found that participants were selective about which ethical issues they documented and were generally hesitant to give full autonomy to the models they developed. 
Piorkowski et al.~\cite{piorkowski2020towards} proposed nine quality dimensions from prior work on software documentation quality analyses and evaluated their usefulness in identifying the quality of AI documentation through a survey.


\subsection{Model Documentation Generation}
Several recent studies have focused on supporting the creation of model documentation. Fang et al.~\cite{fang2020introducing} introduced the Model Card Toolkit (MCT), a set of tools designed to help developers compile and organize the information required for model cards. The toolkit also aids in creating user-friendly interfaces tailored to different audiences. They developed a Python library that can automatically generate model card content based on a predefined JSON schema, using metadata stored in ML Metadata~\cite{mlmetadata}. Similarly, Bhat et al.~\cite{bhat2023aspirations} highlighted the need for better tool support in creating model cards and developed DocML, a tool to help data scientists generate key sections of model documentation more easily in the computational notebook environment.

Some studies presented other ways of representing model-related information. Richards et al.~\cite{richards2020methodology} described a user-centered methodology for creating a specific form of AI documentation, named FactSheets~\cite{arnold2019factsheets}. In their 7-step manual design principles, they proposed a set of questions in each step for different stakeholders to identify the information to be included in the FactSheet and progressively build it. Tsay et al.~\cite{tsay2020aimmx} introduced AIMMX, an Artificial Intelligence Model Metadata Extractor, which automatically extracts high-level contextual information from model repositories. Their goal was to address common challenges in documenting AI models, such as the reliance on manual effort and the absence of standardized documentation practices, which often result in inconsistencies and missing information. They also built a searchable catalog using their metadata extractor to support scalable model discovery and management. Crisan et al.~\cite{crisan2022interactive} introduced and explored the concept of the Interactive Model Card (IMC) as a more accessible and informative way to document deep learning models, with input from ML/AI experts and non-experts. They conducted semi-structured interviews using a think-aloud protocol with 10 ML and AI experts to gather information and feedback on the design of the IMC and implemented a functional prototype of an IMC based on it. They further performed an evaluative study of the functional prototype with 20 non-expert analysts to test the usability and effectiveness of IMC in comparison to traditional standard-model cards.

All these studies operate in a limited setup, such as relying on metadata stored in ML Metadata, using computational notebooks, or focusing on alternative documentation formats. In contrast, our study aims to develop a more widely adoptable approach that aligns with popular and commonly used documentation formats.

%% file: conclusion.tex
\section{Conclusion}
\label{sec:conclusion}

In this paper, we proposed an automated approach for generating model card documentation that follows a standard template, leveraging resources
that are often naturally generated during the model development process, such
as model repository files or academic papers. 
First, we proposed and systematically evaluated \reorganizer, an LLM-based reorganization approach for fitting the contents of a model card into a standard template, thereby improving its structure and clarity. By applying \reorganizer to 48 model cards from Hugging Face, we demonstrated that our LLM-based approach effectively retains most of the original content (a median of 93.8\% of the checklist items), and places it mostly in the correct section (with only 1.9\% of (sub)sections of all the 48 model cards having misplaced contents). Additionally, hallucinated content was rare, affecting only 10.1\% of (sub)sections from all the 48 model cards, and typically appearing in non-technical areas. We also found that \reorganizer{}'s outputs were stable across multiple runs, with a median semantic similarity score of 0.97 for the full documents and at least 0.90 for 87.5\% of the (sub)sections. These results collectively highlight the feasibility of using \reorganizer to automate model card standardization at scale, with minimal human oversight. To support future research and practical adoption, we released \reorganizer and the 48 reorganized and manually curated model cards in a public replication package~\cite{replication-package}.

Beyond reorganization, we introduced \generator, an LLM-based model card generator, to automatically generate new model cards from repository data. Using the reorganized, validated dataset as ground truth, \generator achieved high semantic similarity to the reference content at the document level (mean around 0.9) and produced substantial portions of content with factual correctness. Specifically, 54.17\% of generated model cards were entirely correct, and most others contained only a small number of minor inaccuracies. The remaining inaccuracies mainly stemmed from information mismatches between cited content and generated content, or speculative reasoning, while the absence or variation of some input files (especially associated papers) strongly influenced generation quality. Section-level similarity was lower for sections that require interpretation or synthesis (roughly 0.76–0.80) than for more data-driven sections (e.g., \texttt{Citation details}, \texttt{Training data}, and \texttt{Intersectional results} reached around 0.88–0.92). Ablation analyses showed that regenerating without the associated paper reduced document-level similarity to about 0.91, whereas removing configuration or tokenizer files reduced similarity to about 0.95–0.96, indicating that papers often serve as the primary grounding source. These results demonstrate the potential of \generator to support scalable model card generation, while underscoring the need for high-quality source materials. To foster reproducibility, we released \generator resources alongside the reorganized data in the replication package~\cite{replication-package}.

Together, our findings indicate that LLM-based reorganization and generation can enable scalable, low-effort creation and standardization of model card documentation, with minimal human-in-the-loop requirements.


%% file: appendix.tex
\appendix
\label{sec:appendix}

\section{Model Card Sections' Description for LLM}
\label{sec:section_description}

\begin{lstlisting}[
    caption={Description of the sections of the model card template proposed by Mitchell et al., including additional sections introduced by Toma et al., to inform LLMs about the template structure. Sections introduced by Toma et al. are highlighted in yellow.},
]
\end{lstlisting}
\begin{framed}%
\input{./fig/section_description.md.tex}
\end{framed}

%% file: fig/section_description.md.tex
\subsection*{Model Details}\label{model-details}

This section provides fundamental information about the model, helping
stakeholders understand its context and key characteristics.

\subsubsection*{Person or organization developing
model:}\label{person-or-organization-developing-model}

Describe the individual(s) or organization responsible for the
development of the model. If available, provide background information,
such as their expertise, affiliations, or previous projects they've
worked on, include links to their profiles or official websites for
credibility. This information helps stakeholders identify the creators
and assess credibility.

\subsubsection*{Model date:}\label{model-date}

Specify the timeline of the model's development. Mention key milestones,
such as when the development began, significant updates, and the final
release date. This helps users understand the time frame of the
methodologies and datasets used.

\subsubsection*{Model version:}\label{model-version}

Specify the version of the model, and explain how it differs from other
versions. For example, describe improvements, bug fixes, additional
features, or architectural changes. This aids in tracking updates and
assessing progress. If available, explain how the model differs from
other similar models.

\subsubsection*{Model type:}\label{model-type}

Include every detail about the type of the model, including architecture
details with available explanations (e.g., Transformer, Convolutional
Neural Network) and the specific category it belongs to, such as text
generation, image classification, or reinforcement learning. Highlight
its core components and how they work together to achieve its
functionality. Also, include the model's size and supported context
length if available.

\subsubsection*{Training details:}\label{training-details}

Provide detail with explanation about the training process, including
algorithms used (e.g., supervised learning, reinforcement learning), key
parameters and hyperparameters (e.g., learning rate, number of layers),
fairness constraints or optimization techniques or any other applied
methodologies. Provide enough depth for the reader to understand how the
model achieved its current state.

\subsubsection*{Paper or other resource for more
information:}\label{paper-or-other-resource-for-more-information}

Include links to related papers, repositories, technical blogs,
documentation or other resources that elaborate on the model. If
available, briefly summarize the content of these resources and their
relevance to understanding the model.

\subsubsection*{Citation details:}\label{citation-details}

Provide citation formats (e.g., BibTeX) for referencing the model in
academic or professional work. This allows users to properly acknowledge
the creators.

\subsubsection*{License:}\label{license}

Include all available detail related to the license of the model's
usage. Outline what users can and cannot do with the model, highlighting
any restrictions. If available, include links to the full license text.

\subsubsection*{Contact:}\label{contact}

Provide a contact email or other communication channels for users to ask
questions, report issues or provide feedback. If available, include
additional resources like forums or FAQs.

\begin{center}\rule{0.5\linewidth}{0.5pt}\end{center}

\subsection*{Intended Use}\label{intended-use}

This section outlines the intended applications of the model.

\subsubsection*{Primary intended uses:}\label{primary-intended-uses}

Describe the primary purposes for which the model was created. State
whether it was tailored for specific tasks (e.g., sentiment analysis) or
built as a general-purpose tool. Be specific about tasks or domains.
Explain the capabilities of the model. Include examples of how the model
can be applied in real-world scenarios. If available, explain the
input-output structure of the model.

\subsubsection*{Primary intended users:}\label{primary-intended-users}

Identify the target audience for the model. Examples include
researchers, developers, businesses, or educators. Describe their
expected level of expertise and typical use cases.

\subsubsection*{Out-of-scope uses:}\label{out-of-scope-uses}

List applications the model is not designed for, including potential
misuse cases. Highlight situations where the model might be misapplied.
Provide examples of related technologies or contexts that might lead to
confusion. Suggest alternative models that are more suitable for those
contexts if applicable.

\begin{center}\rule{0.5\linewidth}{0.5pt}\end{center}

\begin{mdframed}[backgroundcolor=yellow!20, hidealllines=true]
\subsection*{How to Use}\label{how-to-use}

This section outlines how to use the model.

Include all details on model usage and its example outputs, including
input-output structure, settings, code snippets, explanations, and
sample input-outputs. If available, add links to documentation and
tutorials for further guidance.
\end{mdframed}

\begin{center}\rule{0.5\linewidth}{0.5pt}\end{center}

\subsection*{Factors}\label{factors}

This section addresses variables that may impact the model's
performance.

\subsubsection*{Relevant factors:}\label{relevant-factors}

List key factors that influence the model's performance, such as
demographic variations, environmental conditions, or data collection
methods. Explain how these factors were identified and why they are
important.

\subsubsection*{Evaluation factors:}\label{evaluation-factors}

Indicate which factors are analyzed and reported during model
evaluation. If these differ from the relevant factors, explain why they
were selected. For instance, you might focus on accuracy metrics over
demographic fairness for a specific evaluation.

\begin{center}\rule{0.5\linewidth}{0.5pt}\end{center}

\subsection*{Metrics}\label{metrics}

This section describes how the model's performance is evaluated.

\subsubsection*{Model performance
measures:}\label{model-performance-measures}

Discuss the metrics used to assess the model's effectiveness (e.g.,
accuracy, F1 score, precision, recall). Justify the selection of these
metrics and why they are more suitable than others.

\subsubsection*{Decision thresholds:}\label{decision-thresholds}

Describe thresholds used in decision-making (e.g., classifying spam
emails) and their rationale. Include any empirical evidence supporting
these decisions.

\subsubsection*{Variation approaches:}\label{variation-approaches}

Explain how performance metrics were calculated or estimated, including
any uncertainty measures. For example, describe cross-validation,
bootstrapping, or other statistical methods used to ensure robust
measurements.

\begin{center}\rule{0.5\linewidth}{0.5pt}\end{center}

\subsection*{Evaluation Data}\label{evaluation-data}

This section provides details about the datasets used to evaluate the
model.

\subsubsection*{Datasets:}\label{datasets}

Provide information on the datasets used to evaluate the model. Include
every details available about the data like size, diversity, source,
public availability or proprietary.

\subsubsection*{Motivation:}\label{motivation}

Explain why these datasets were chosen for evaluation. Discuss their
relevance to the model's intended use and how well they represent
real-world scenarios.

\subsubsection*{Preprocessing:}\label{preprocessing}

Describe the preprocessing steps applied to the evaluation data in
detail with explanation. Examples include normalization, encoding, or
filtering. If available, explain how these steps align with the model's
design and intended use.

\begin{center}\rule{0.5\linewidth}{0.5pt}\end{center}

\subsection*{Training Data}\label{training-data}

This section provides details about the datasets used to train the
model.

\subsubsection*{Datasets:}\label{datasets-1}

Provide information on the datasets used to train the model. Include
every details available about the data like size, structure, features,
diversity. If the data is publicly available, include links.

\subsubsection*{Motivation:}\label{motivation-1}

Justify the choice of datasets for training, highlighting their
suitability for the model's purpose and intended applications.

\subsubsection*{Preprocessing:}\label{preprocessing-1}

Describe the preprocessing steps applied to the training data in detail
with explanation. Include examples like text tokenization, image
resizing, or outlier removal. Explain how these steps improved training
efficiency or accuracy.

\begin{center}\rule{0.5\linewidth}{0.5pt}\end{center}

\subsection*{Quantitative Analyses}\label{quantitative-analyses}

This section presents disaggregated evaluation results.

\subsubsection*{Unitary results:}\label{unitary-results}

Present performance results for each individual factors identified in
the Factors section. For example, show accuracy rates for different
demographic groups or under varying environmental conditions.

\subsubsection*{Intersectional results:}\label{intersectional-results}

Present performance results across combinations of factors. For
instance, analyze accuracy for a specific demographic group within a
particular geographic location.

\begin{center}\rule{0.5\linewidth}{0.5pt}\end{center}

\begin{mdframed}[backgroundcolor=yellow!20, hidealllines=true]
\subsection*{Memory or Hardware Requirements}\label{memory-or-hardware-requirements}

This section outlines the memory or hardware requirements for loading,
deploying, and training the model.

\subsubsection*{Loading Requirements:}\label{loading-requirements}

If available, specify the memory and hardware requirements (e.g.,
RAM/VRAM size, disk space, CPU/GPU/TPU) to load the model.

\subsubsection*{Deploying Requirements:}\label{deploying-requirements}

If available, specify the memory or hardware requirements to run and
serve the model.

\subsubsection*{Training or Fine-tuning
Requirements:}\label{training-or-fine-tuning-requirements}

If available, specify the memory or hardware requirements to train or
finetune the model.
\end{mdframed}

\begin{center}\rule{0.5\linewidth}{0.5pt}\end{center}

\subsection*{Ethical Considerations}\label{ethical-considerations}

This section discusses the ethical considerations in model development,
including challenges, risks, and solutions.

Specify if sensitive data (e.g., personal information, protected
attributes) was used. Identify potential risks associated with the
model's application, their likelihood, and severity, especially in
critical areas like healthcare or public safety. Describe risk
mitigation strategies used during development and risks in model usage,
including potential harm and affected groups. If risks are unknown, note
that they were considered. Highlight the known model use cases that are
especially fraught. Highlight efforts to address these challenges and
acknowledge areas requiring further exploration.

\begin{center}\rule{0.5\linewidth}{0.5pt}\end{center}

\subsection*{Caveats and
Recommendations}\label{caveats-and-recommendations}

This section lists unresolved issues and provides guidance for users.

\subsubsection*{Caveats:}\label{caveats}

List any limitations or areas of concern not addressed earlier. For
example: gaps in evaluation datasets (e.g., missing demographic groups),
suggestions for future testing or research, ideal characteristics of
datasets for further evaluation etc.

\subsubsection*{Recommendations:}\label{recommendations}

Suggest best practices for using the model and areas for further
testing. Provide actionable recommendations for users to maximize the
model's benefits while mitigating risks.

\begin{center}\rule{0.5\linewidth}{0.5pt}\end{center}

%% file: bibtex.bib
@article{richards2020methodology,
  title={A methodology for creating AI FactSheets},
  author={Richards, John and Piorkowski, David and Hind, Michael and Houde, Stephanie and Mojsilovi{\'c}, Aleksandra},
  journal={arXiv preprint arXiv:2006.13796},
  year={2020}
}

@article{richards2021human,
  title={A Human-Centered Methodology for Creating AI FactSheets.},
  author={Richards, John T and Piorkowski, David and Hind, Michael and Houde, Stephanie and Mojsilovic, Aleksandra and Varshney, Kush R},
  journal={IEEE Data Eng. Bull.},
  volume={44},
  number={4},
  pages={47--58},
  year={2021}
}

@article{fang2020introducing,
  title={Introducing the model card toolkit for easier model transparency reporting},
  author={Fang, Huanming and Miao, Hui and Shukla, Karan and Nanas, Dan and Xu, Catherina and Greer, Christina and Polyzotis, Neoklis and Doshi, Tulsee and Deng, Tiffany and Mitchell, Margaret and others},
  journal={Google AI Blog},
  year={2020}
}

@article{piorkowski2020towards,
  title={Towards evaluating and eliciting high-quality documentation for intelligent systems},
  author={Piorkowski, David and Gonz{\'a}lez, Daniel and Richards, John and Houde, Stephanie},
  journal={arXiv preprint arXiv:2011.08774},
  year={2020}
}

@misc {luca_papariello_2024,
  author={{Luca Papariello}},
  title={xlm-roberta-base-language-detection (Revision 9865598)},
  year=2024,
  url={https://huggingface.co/papluca/xlm-roberta-base-language-detection},
  doi={10.57967/hf/2064},
  publisher={Hugging Face}
}

@inproceedings{mitchell2019model,
  title={{Model Cards for Model Reporting}},
  author={Mitchell, Margaret and Wu, Simone and Zaldivar, Andrew and Barnes, Parker and Vasserman, Lucy and Hutchinson, Ben and Spitzer, Elena and Raji, Inioluwa Deborah and Gebru, Timnit},
  booktitle={Proceedings of the conference on fairness, accountability, and transparency},
  pages={220--229},
  year={2019}
}

@article{liang2024s,
  title={{What's documented in AI? Systematic Analysis of 32K AI Model Cards}},
  author={Liang, Weixin and Rajani, Nazneen and Yang, Xinyu and Ozoani, Ezinwanne and Wu, Eric and Chen, Yiqun and Smith, Daniel Scott and Zou, James},
  journal={arXiv preprint arXiv:2402.05160},
  year={2024}
}

@inproceedings{bhat2023aspirations,
  title={{Aspirations and Practice of ML Model Documentation: Moving the Needle with Nudging and Traceability}},
  author={Bhat, Avinash and Coursey, Austin and Hu, Grace and Li, Sixian and Nahar, Nadia and Zhou, Shurui and K{\"a}stner, Christian and Guo, Jin LC},
  booktitle={Proceedings of the 2023 CHI Conference on Human Factors in Computing Systems},
  pages={1--17},
  year={2023}
}

@article{oreamuno2024state,
  title={{The State of Documentation Practices of Third-party Machine Learning Models and Datasets}},
  author={Oreamuno, Ernesto Lang and Khan, Rohan Faiyaz and Bangash, Abdul Ali and Stinson, Catherine and Adams, Bram},
  journal={IEEE Software},
  year={2024},
  publisher={IEEE}
}

@article{nunes2024using,
  title={{Using Model Cards for ethical reflection on machine learning models: an interview-based study}},
  author={Nunes, Jos{\'e} Luiz and Barbosa, Gabriel DJ and de Souza, Clarisse Sieckenius and Barbosa, Simone DJ},
  journal={Journal on Interactive Systems},
  volume={15},
  number={1},
  pages={1--19},
  year={2024}
}

@article{cohan2020specter,
  title={Specter: Document-level representation learning using citation-informed transformers},
  author={Cohan, Arman and Feldman, Sergey and Beltagy, Iz and Downey, Doug and Weld, Daniel S},
  journal={arXiv preprint arXiv:2004.07180},
  year={2020}
}

@inproceedings{toma2025answering,
  title={Answering User Questions about Machine Learning Models through Standardized Model Cards},
  author={Toma, Tajkia Rahman and Grewal, Balreet and Bezemer, Cor-Paul},
  booktitle={2025 IEEE/ACM 47th International Conference on Software Engineering (ICSE)},
  pages={603--603},
  year={2025},
  organization={IEEE Computer Society}
}

@article{schober2018correlation,
  title={Correlation coefficients: appropriate use and interpretation},
  author={Schober, Patrick and Boer, Christa and Schwarte, Lothar A},
  journal={Anesthesia \& analgesia},
  volume={126},
  number={5},
  pages={1763--1768},
  year={2018},
  publisher={LWW}
}

@article{abbas2023relationship,
  title={On the relationship between similar requirements and similar software: A case study in the railway domain},
  author={Abbas, Muhammad and Ferrari, Alessio and Shatnawi, Anas and Enoiu, Eduard and Saadatmand, Mehrdad and Sundmark, Daniel},
  journal={Requirements Engineering},
  volume={28},
  number={1},
  pages={23--47},
  year={2023},
  publisher={Springer}
}

@inproceedings{zainyte2021challenges,
  title={{Challenges and future directions for accountable machine learning}},
  author={Zainyte, Agne and Pang, Wei},
  booktitle={CEUR Workshop Proceedings},
  volume={2894},
  pages={40--47},
  year={2021},
  organization={CEUR-WS},
  issn={1613-0073}
}

@inproceedings{taraghi2024deep,
  title={Deep learning model reuse in the huggingface community: Challenges, benefit and trends},
  author={Taraghi, Mina and Dorcelus, Gianolli and Foundjem, Armstrong and Tambon, Florian and Khomh, Foutse},
  booktitle={2024 IEEE International Conference on Software Analysis, Evolution and Reengineering (SANER)},
  pages={512--523},
  year={2024},
  organization={IEEE}
}

@inproceedings{tsay2020aimmx,
  title={{AIMMX: Artificial Intelligence Model Metadata Extractor}},
  author={Tsay, Jason and Braz, Alan and Hirzel, Martin and Shinnar, Avraham and Mummert, Todd},
  booktitle={Proceedings of the 17th International Conference on Mining Software Repositories},
  pages={81--92},
  year={2020}
}

@misc{mlmetadata,
  title={ML Metadata},
  author={{Google}},
  year={2019},
  howpublished={\url{https://github.com/google/ml-metadata}},
  note={Accessed: 2025-04-06}
}

@inproceedings{crisan2022interactive,
  title={{Interactive Model Cards: A Human-Centered Approach to Model Documentation}},
  author={Crisan, Anamaria and Drouhard, Margaret and Vig, Jesse and Rajani, Nazneen},
  booktitle={Proceedings of the 2022 ACM Conference on Fairness, Accountability, and Transparency},
  pages={427--439},
  year={2022},
  isbn={9781450393522},
  publisher={Association for Computing Machinery},
  address={New York, NY, USA},
  url={https://doi.org/10.1145/3531146.3533108},
  doi={10.1145/3531146.3533108},
  numpages={13},
  location={Seoul, Republic of Korea},
  series={FAccT '22}
}

@article{verga2024replacing,
  title={Replacing judges with juries: Evaluating LLM generations with a panel of diverse models},
  author={Verga, Pat and Hofstatter, Sebastian and Althammer, Sophia and Su, Yixuan and Piktus, Aleksandra and Arkhangorodsky, Arkady and Xu, Minjie and White, Naomi and Lewis, Patrick},
  journal={arXiv preprint arXiv:2404.18796},
  year={2024}
}

@article{li2024software,
  title={Software Engineering and Foundation Models: Insights from Industry Blogs Using a Jury of Foundation Models},
  author={Li, Hao and Bezemer, Cor-Paul and Hassan, Ahmed E},
  journal={arXiv preprint arXiv:2410.09012},
  year={2024}
}

@article{kenton2024scalable,
  title={On scalable oversight with weak llms judging strong llms},
  author={Kenton, Zachary and Siegel, Noah and Kram{\'a}r, J{\'a}nos and Brown-Cohen, Jonah and Albanie, Samuel and Bulian, Jannis and Agarwal, Rishabh and Lindner, David and Tang, Yunhao and Goodman, Noah and others},
  journal={Advances in Neural Information Processing Systems},
  volume={37},
  pages={75229--75276},
  year={2024}
}

@article{seo2025large,
  title={Large Language Models as Evaluators in Education: Verification of Feedback Consistency and Accuracy.},
  author={Seo, Hyein and Hwang, Taewook and Jung, Jeesu and Kang, Hyeonseok and Namgoong, Hyuk and Lee, Yohan and Jung, Sangkeun},
  journal={Applied Sciences (2076-3417)},
  volume={15},
  number={2},
  year={2025}
}

@misc{chiang2024chatbot,
  title={Chatbot Arena: An Open Platform for Evaluating LLMs by Human Preference},
  author={Wei-Lin Chiang and Lianmin Zheng and Ying Sheng and Anastasios Nikolas Angelopoulos and Tianle Li and Dacheng Li and Hao Zhang and Banghua Zhu and Michael Jordan and Joseph E. Gonzalez and Ion Stoica},
  year={2024},
  eprint={2403.04132},
  archivePrefix={arXiv},
  primaryClass={cs.AI}
}

@inproceedings{pepe2024hugging,
  title={{How do Hugging Face Models Document Datasets, Bias, and Licenses? An Empirical Study}},
  author={Pepe, Federica and Nardone, Vittoria and Mastropaolo, Antonio and Bavota, Gabriele and Canfora, Gerardo and Di Penta, Massimiliano},
  booktitle={Proceedings of the 32nd IEEE/ACM International Conference on Program Comprehension},
  pages={370--381},
  year={2024}
}

@inproceedings{gao2024documenting,
  title={Documenting ethical considerations in open source ai models},
  author={Gao, Haoyu and Zahedi, Mansooreh and Treude, Christoph and Rosenstock, Sarita and Cheong, Marc},
  booktitle={Proceedings of the 18th ACM/IEEE International Symposium on Empirical Software Engineering and Measurement},
  pages={177--188},
  year={2024}
}

@inproceedings{nunes2022using,
  title={Using model cards for ethical reflection: a qualitative exploration},
  author={Nunes, Jos{\'e} Luiz and Barbosa, Gabriel DJ and De Souza, Clarisse Sieckenius and Lopes, Helio and Barbosa, Simone DJ},
  booktitle={Proceedings of the 21st Brazilian Symposium on Human Factors in Computing Systems},
  pages={1--11},
  year={2022}
}

@article{arnold2019factsheets,
  title={{FactSheets: Increasing trust in AI services through supplier's declarations of conformity}},
  author={Arnold, Matthew and Bellamy, Rachel KE and Hind, Michael and Houde, Stephanie and Mehta, Sameep and Mojsilovi{\'c}, Aleksandra and Nair, Ravi and Ramamurthy, K Natesan and Olteanu, Alexandra and Piorkowski, David and others},
  journal={IBM Journal of Research and Development},
  volume={63},
  number={4/5},
  pages={6--1},
  year={2019},
  publisher={IBM}
}

@article{zheng2023progressive,
  title={Progressive-hint prompting improves reasoning in large language models},
  author={Zheng, Chuanyang and Liu, Zhengying and Xie, Enze and Li, Zhenguo and Li, Yu},
  journal={arXiv preprint arXiv:2304.09797},
  year={2023}
}

@article{krishna2024understanding,
  title={Understanding the effects of iterative prompting on truthfulness},
  author={Krishna, Satyapriya and Agarwal, Chirag and Lakkaraju, Himabindu},
  journal={arXiv preprint arXiv:2402.06625},
  year={2024}
}

@article{chang2024efficient,
  title={Efficient prompting methods for large language models: A survey},
  author={Chang, Kaiyan and Xu, Songcheng and Wang, Chenglong and Luo, Yingfeng and Liu, Xiaoqian and Xiao, Tong and Zhu, Jingbo},
  journal={arXiv preprint arXiv:2404.01077},
  year={2024}
}

@misc{replication-package,
  title={{Replication Package}},
  author={Toma, Tajkia Rahman and Grewal, Balreet and Bezemer, Cor-Paul},
  howpublished={\url{https://github.com/asgaardlab/mc-reorganization-and-generation}},
  year={2026}
}

@article{latendresse2024exploratory,
  author={Latendresse, Jasmine and Abedu, Samuel and Abdellatif, Ahmad and Shihab, Emad},
  title={An Exploratory Study on Machine Learning Model Management},
  year={2024},
  issue_date={January 2025},
  publisher={Association for Computing Machinery},
  address={New York, NY, USA},
  volume={34},
  number={1},
  issn={1049-331X},
  url={https://doi.org/10.1145/3688841},
  doi={10.1145/3688841},
  journal={ACM Trans. Softw. Eng. Methodol.},
  month=dec,
  articleno={16},
  numpages={31},
}

@misc{huggingfaceAppendix,
  author={Ozoani, Ezi and Gerchick, Marissa and Mitchell, Margaret},
  title={{A}ppendix --- huggingface.co},
  howpublished={\url{https://huggingface.co/docs/hub/model-card-appendix\#what-do-you-dislike-about-model-cards}},
  year={2022},
  note={[Accessed 04-04-2025]},
}

@article{habehh2021machine,
  title={Machine learning in healthcare},
  author={Habehh, Hafsa and Gohel, Suril},
  journal={Current genomics},
  volume={22},
  number={4},
  pages={291--300},
  year={2021},
  publisher={Bentham Science Publishers direct}
}

@article{ahmed2022artificial,
  title={Artificial intelligence and machine learning in finance: A bibliometric review},
  author={Ahmed, Shamima and Alshater, Muneer M and El Ammari, Anis and Hammami, Helmi},
  journal={Research in International Business and Finance},
  volume={61},
  pages={101646},
  year={2022},
  publisher={Elsevier}
}

@article{goodell2021artificial,
  title={Artificial intelligence and machine learning in finance: Identifying foundations, themes, and research clusters from bibliometric analysis},
  author={Goodell, John W and Kumar, Satish and Lim, Weng Marc and Pattnaik, Debidutta},
  journal={Journal of Behavioral and Experimental Finance},
  volume={32},
  pages={100577},
  year={2021},
  publisher={Elsevier}
}

@article{travaini2022machine,
  title={Machine learning and criminal justice: A systematic review of advanced methodology for recidivism risk prediction},
  author={Travaini, Guido Vittorio and Pacchioni, Federico and Bellumore, Silvia and Bosia, Marta and De Micco, Francesco},
  journal={International journal of environmental research and public health},
  volume={19},
  number={17},
  pages={10594},
  year={2022},
  publisher={MDPI}
}

@article{zavrvsnik2021algorithmic,
  title={Algorithmic justice: Algorithms and big data in criminal justice settings},
  author={Zavr{\v{s}}nik, Ale{\v{s}}},
  journal={European Journal of criminology},
  volume={18},
  number={5},
  pages={623--642},
  year={2021},
  publisher={SAGE Publications Sage UK: London, England}
}

@article{javaid2022significance,
  title={Significance of machine learning in healthcare: Features, pillars and applications},
  author={Javaid, Mohd and Haleem, Abid and Singh, Ravi Pratap and Suman, Rajiv and Rab, Shanay},
  journal={International Journal of Intelligent Networks},
  volume={3},
  pages={58--73},
  year={2022},
  publisher={Elsevier}
}
